\documentclass[11pt]{article} 
\usepackage{jheppub}

\usepackage{amsmath,amsfonts,amsbsy,amssymb,array,accents,dsfont}
\usepackage{enumerate}
\usepackage[shortlabels]{enumitem}
\usepackage{graphicx} 
\usepackage{subcaption} 
\usepackage{upgreek}
\usepackage{bm}
\usepackage{slashed}

\let\includefigures=\iftrue
\let\useblackboard==\iftrue
\newfam\black

\usepackage[dvipsames]{xcolor}

\definecolor{myblue}{RGB}{85,130,255}
\definecolor{myred}{RGB}{200, 45, 40}

\PassOptionsToPackage{ 
     colorlinks=true,
     linkcolor=myBlue,
     urlcolor=blue,
     filecolor=black,
     citecolor=myRed,
}{hyperref}

\usepackage{xparse}
\usepackage{xspace}

\NewDocumentCommand\eqn{om}{%
  \IfNoValueTF{#1}
     {\[ #2 \]}
     {\begin{equation}\label{#1} #2  \end{equation} \expandafter\newcommand\csname #1\endcsname{\eqref{#1}\xspace}\ignorespaces}
}
\NewDocumentCommand\eqna{om}{%
  \IfNoValueTF{#1}
    {\begin{align*} #2 \end{align*}}
    {\begin{equation}\label{#1}\begin{split} #2  \end{split}\end{equation} \expandafter\def\csname #1\endcsname{\eqref{#1}\xspace}\ignorespaces}
}

\newcommand{\rcite}{\cite}

\def\ytil{{\tilde y}}

\def\Re{\it Re}
\def\Im{\it Im}

\def\tauperp{\tau_{\sst \!\perp}^{~}}
\def\taupar{\tau_{\sst \parallel}^{~}}

\def\asymlabel{{\ell}}
\def\asym{{(\ell)}}
\def\asymtil{{(\tilde\ell)}}
\def\I{{\sfm}}

\def\llangle{{\big\langle\hskip -2.8pt\big\langle}}
\def\rrangle{{\big\rangle\hskip -2.8pt\big\rangle}}
\def\LLangle{{\Big\langle\hskip -4pt\Big\langle}}
\def\RRangle{{\Big\rangle\hskip -4pt\Big\rangle}}

\def\sl{\text{sl}}

\def\vareps{\varepsilon}

\def\str{{\rm str}}

\def\sltwo{\ensuremath{SL(2,\bR)}}

\def\sutwo{{SU(2)}}

\def\uone{U(1)}

\def\NSfive{{\sst\rm NS5}}

\def\tight#1{\! #1 \!}  

\def\({\left(}
\def\){\right)}
\def\[{\left[}
\def\]{\right]}

\def\ie{{i.e.}}
\def\eg{{e.g.}}

\def\etc{{etc}}

\def\apl{\alpha'_{\textit{little}}}

\def\th{{\rm th}}

\def\nfive{{n_5}}

\def\none{{n_1}}

\def\sfA{{\mathsf A}}
\def\sfB{{\mathsf B}}

\def\sfF{{\mathsf F}}

\def\sfH{{\mathsf H}}

\def\sfX{{\mathsf X}}

\def\sfZ{{\mathsf Z}}
\def\sfa{{\mathsf a}}
\def\sfb{{\mathsf b}}

\def\sfm{{\mathsf m}}

\def\sfu{{\mathsf u}}
\def\sfv{{\mathsf v}}

\def\sfx{{\mathsf x}}
\def\sfy{{\mathsf y}}

\def\mfa{{\mathfrak a}}

\DeclareMathSymbol{\medhatsym}{\mathord}{largesymbols}{"62} 

\DeclareMathSymbol{\medtildesym}{\mathord}{largesymbols}{"65}

\makeatletter
\newcommand*\rel@kern[1]{\kern#1\dimexpr\macc@kerna}
\newcommand*\widebar[1]{%
  \begingroup
  \def\mathaccent##1##2{%
    \rel@kern{0.8}%
    \overline{\rel@kern{-0.8}\macc@nucleus\rel@kern{0.2}}%
    \rel@kern{-0.2}%
  }%
  \macc@depth\@ne
  \let\math@bgroup\@empty \let\math@egroup\macc@set@skewchar
  \mathsurround\z@ \frozen@everymath{\mathgroup\macc@group\relax}%
  \macc@set@skewchar\relax
  \let\mathaccentV\macc@nested@a
  \macc@nested@a\relax111{#1}%
  \endgroup
}
\makeatother

\def\ytil{{\tilde y}}

\def\half{\frac12}

\def\One{{\hbox{1\kern-1mm l}}}

\def\barray{\begin{array}}
\def\earray{\end{array}}
\def\be{\begin{equation}}
\def\ee{\end{equation}}
\def\bea{\begin{align}}
\def\eea{\end{align}}
\def\bal{\begin{align}}
\def\eal{\end{align}}
\def\nn{\nonumber}

\newcommand{\bR}{{\mathbb R}}
\newcommand{\bS}{{\mathbb S}}
\newcommand{\bT}{{\mathbb T}}

\newcommand{\bZ}{{\mathbb Z}}

\definecolor{cardinal}{rgb}{0.6,0,0}
\definecolor{darkgreen}{rgb}{0,0.4,0}
\definecolor{green}{rgb}{0,0.4,0}
\definecolor{golden}{rgb}{0.92, 0.7, 0}
\definecolor{midnight}{rgb}{0, 0, 0.5}
\definecolor{darkblue}{rgb}{0, 0, 0.7}

\numberwithin{equation}{section}

\mathchardef\mhyphen="2D

  \def\cF{\mathcal {F}}
\def\cG{\mathcal {G}} \def\cH{\mathcal {H}} \def\cI{\mathcal {I}}
  
\def\cM{\mathcal {M}}  
  
\def\cS{\mathcal {S}}  
 \def\cW{\mathcal {W}}

\def\one{{\hbox{\kern+.5mm 1\kern-.8mm l}}}
\def\zero{{\hbox{0\kern-1.5mm 0}}}

\def\id{\textrm{id}}

\def\id{{1 \kern-.28em {\rm l}}}

\def\journal#1&#2(#3){\unskip, \sl #1\ \bf #2 \rm(19#3) }
\def\andjournal#1&#2(#3){\sl #1~\bf #2 \rm (19#3) }

\def\ie{{\it i.e.}}
\def\eg{{\it e.g.}}

\def\etc{{\it etc}}

\def\sst{\scriptscriptstyle}

\def\half{\frac12}

\def\One{{1\hskip -3pt {\rm l}}}

\catcode`\@=11
\def\slash#1{\mathord{\mathpalette\c@ncel{#1}}}
\def\vareps{\varepsilon}
\def\underrel#1\over#2{\mathrel{\mathop{\kern\z@#1}\limits_{#2}}}

\catcode`\@=12

\def\exp{{\rm exp}}

\def\ie{{\it i.e.}}
\def\eg{{\it e.g.}}

\title{
{
BPS Fivebrane Stars III:
Effective Actions
}}

\author{Emil J. Martinec$^a$, Yoav Zigdon$^{b}$\\}

\affiliation[a]{
Kadanoff Center for Theoretical Physics, Enrico Fermi Institute, and Department of Physics\\ 
University of Chicago\\ 
5640 S. Ellis Ave.\\
Chicago IL 60637  USA\\ 
}

\affiliation[b]{
Department of Applied Mathematics and Theoretical Physics\\
University of Cambridge\\
Cambridge, CB3 0WA, United Kingdom \\
}

 \emailAdd{e-martinec@uchicago.edu}
 \emailAdd{ yz910@cam.ac.uk}

\abstract{
An effective action for NS5-branes coupled to supergravity is used to derive the full 10d form of horizon-free BPS solutions of fivebranes carrying momentum waves, including both transverse scalar and internal gauge excitations of the branes. 
When internal modes are highly excited, we find solutions that plausibly mediate the transition between the Coulomb phase of NS5-branes and the black hole phase. 
We also compute the two-point functions of fivebrane density fluctuations and of gravitons absorbed by the branes. Finally, we begin an exploration of near-BPS perturbations of the fivebrane ensemble, and propose the use of the brane+bulk effective action as a tool to explore the black hole phase, even in the AdS decoupling limit.
}
\begin{document}
\maketitle
\hypersetup{pageanchor=false}
\hypersetup{pageanchor=true}
\pagenumbering{arabic}





\vskip 2cm


\section{Introduction} 
\label{sec:intro}

Neveu-Schwarz fivebranes are solitons of closed string dynamics~-- heavy objects sourcing a semiclassical background.  They have played an important role in the study of black holes in string theory~\rcite{Strominger:1996sh}; and they are key ingredients of a rich collection of BPS solutions of string theory, for which the supergravity field equations can be explicitly solved (or at least reduced to a system of linear equations; see for instance~\rcite{Mathur:2005zp,Kanitscheider:2007wq,Skenderis:2008qn,Shigemori:2020yuo,Bena:2022rna} for reviews).  Many of these BPS solutions have parametrically deep throats with large redshifts; they thus have some features in common with black holes, and the aspects by which they differ inform our understanding of the black hole phase.  

In~\rcite{Martinec:2023xvf,Martinec:2023gte} (see also~\rcite{Alday:2006nd,Raju:2018xue}) the properties of generic two-charge BPS states of fivebranes carrying momentum waves were studied through their ensemble average.  
The typical solution can be thought of as a BPS fivebrane star, supported on the fivebranes' Coulomb branch by the momentum excitations of the scalars specifying their transverse location.  Our initial analysis considered only transverse (scalar) excitations of the fivebranes, whose effect is to cause the fivebrane to undergo a random walk in its transverse space.  In this work,
we extend the analysis to include internal excitations of the fivebrane gauge multiplet following~\rcite{Kanitscheider:2007wq}, so that the full set of bosonic fivebrane excitations is involved.  These internal modes take up roughly half of the excitation budget, but don't contribute to supporting the fivebrane against collapse.  Exciting these modes while fixing the charges renders the system smaller because transverse excitations possess less energy. As we show, the internal gauge modes give rise to fivebrane bound states that become more similar to the black hole solution, in that the redshift and the length of the throat increase. 
Thus, one possible route to collapse of the fivebrane star is to favor internal over transverse modes in the distribution of excitations.
Note however that when we make the distribution between internal and external excitations highly lopsided in favor of the internal ones, we are giving up half the entropy, and so this will be thermodynamically unfavorable unless there are compensating dynamical effects.  

In the course of our analysis, we develop 
an effective action approach to fivebrane dynamics.  Initial work on the BPS system proceeded indirectly, employing smeared versions of known two-charge solutions sourced by fundamental strings~\rcite{Callan:1995hn,Dabholkar:1995nc} together with U-duality transformations to find smeared versions of two-charge solutions involving NS5-branes~\rcite{Lunin:2001fv}.  One advantage of the effective action approach is that it allows us to construct the solution directly in the duality frame of interest, without smearing the source, and in addition to exhibit the coupling of the bulk supergravity fields to the fivebrane source.

Because NS5-branes are solitonic, there is a tight correlation between excitations of the branes and excitations of the bulk supergravity fields~-- there is no limit where one can turn off the supergravity back-reaction.%
\footnote{The string coupling scales out of the NS5-brane decoupling limit, as we review presently.}
One sees this feature explicitly in the BPS solutions, 
where the bulk supergravity fields and the gauge supermultiplet on the brane are both expressed in terms of the same set of eight bosonic profile functions (and more generally, their fermionic superpartners~\rcite{Taylor:2005db}).

The effective action formalism is expected to prove useful in the analysis of the near-BPS regime.  For non-BPS solutions, chasing solutions through U-duality transformations is less useful, as the light non-BPS excitations in one duality frame are typically completely different from those in another duality frame.  

This article is organized as follows. In section~\ref{sec:review}, we review properties of supersymmetric NS5-branes in the context of string theory, and provide examples of known solutions of supergravity and worldsheet conformal field theories describing string propagation in the presence of fivebranes. In section~\ref{sec:STsolns}, we utilize the effective action formulation to derive the 1/2-BPS Lunin-Mathur solutions~\cite{Lunin:2001fv} in the NS5-P duality frame, and their extension to include bosonic internal gauge mode excitations, following reference~\cite{Kanitscheider:2007wq}. In section~\ref{sec:NS5-P ensemble}, we perform an ensemble average of the solutions and generalize the previous analysis in~\cite{Martinec:2023xvf} by incorporating the gauge multiplet. We also calculate two-point functions of density peturbations and gravitons absorbed by the NS5-branes. In section~\ref{sec:perts}, we calculate a contribution to the effective action governing near-BPS fluctuations, which we expect to be useful to begin addressing the question of the stability of BPS stars, and provide a discussion about our results and future directions in section~\ref{sec:discussion}. Appendix~\ref{app:1source} argues that in the fivebrane decoupling limit, the transverse and internal excitations are pure gauge for a single brane source.

Among the questions one wishes to address is the stability of these BPS stars~-- do they immediately (if slowly) collapse to a black hole when slightly perturbed, or do they find a nearby stellar equilibrium configuration?  These and other possible extensions of our work are explored in section~\ref{sec:discussion}.

We begin with a review of NS5-branes and their properties in string theory.


\section{Review of BPS fivebranes} 
\label{sec:review}

The background sourced by $\nfive$ coincident, unexcited NS5-branes is given by \cite{Callan:1991at}
\begin{align}
ds^2 &= \sfH_5\big( dr^2 + r^2d\Omega_3^2\big) + dx^2_{\bR^{5,1}}
\nn\\[.2cm]
dB_2 &= *_\perp d\sfH_5
~~,~~~~ 
e^{2\Phi} = g_s^2 \sfH_5
~~,~~~~
\sfH_5 = 1 + \frac{n_5\alpha'}{r^2} ~.
\label{NS5stack}
\end{align}
The decoupling limit 
\be
\label{decoupling}
g_s\to 0
~~,~~~~
e^{2\rho} = \frac{r^2}{g_s^2\alpha'} ~~ {\rm fixed}  
\ee
suppresses the effects of the constant term in~$\sfH_5$, and
yields an exact closed string background~\rcite{Callan:1991at}
\begin{align}
ds^2 &= \nfive\alpha'\big(d\rho^2 + d\Omega_3^2\big) + dx^2_{\bR^{5,1}} 
\nn\\[.2cm]
dB_2 &= -n_5 \alpha' {\rm vol_{\Omega_3}}
~~,~~~~
e^{2\Phi} = n_5 e^{-2\rho} ~,
\label{CHS geom}
\end{align}
whose worldsheet description consists of a free scalar with linear dilaton for the radial direction, times an $\sutwo$ WZW model for the transverse angular directions, together with free field theory for the 5+1 directions along the branes.
Note that the supersymmetric $\sutwo$ WZW model has a minimum level of $\nfive\tight=2$~-- there is no description of the throat of a single fivebrane, only two or more coincident fivebranes.

Unfortunately there is no string S-matrix in this background~-- strings simply propagate freely into the region of large string coupling at large negative $\rho$ and we lose control over the dynamics.

\subsection{Non-singularity of the Coulomb branch of NS5-branes}

Perturbative string theory arises when we separate the fivebranes onto their Coulomb branch~\rcite{Giveon:1999px,Giveon:1999tq}.  The harmonic function $\sfH_5$ characterizing the background~\eqref{NS5stack} becomes (in Cartesian coordinates for the transverse space)
\be
\label{H5coul}
\sfH_5 = 1 + \sum_{\I=1}^\nfive \frac{\alpha'}{|\sfx-\sfx_\I|^2} ~;
\ee
the decoupling limit is now 
\be
\label{decoupling2}
g_s\to 0
~~,~~~~
\hat\sfx  = \frac{\sfx}{g_s\sqrt{\alpha'}} 
~~,~~~
\hat\sfx_\I  = \frac{\sfx_\I}{g_s\sqrt{\alpha'}}
~~ {\rm fixed} ~.
\ee
Placing them in a $\bZ_\nfive$-symmetric array in a transverse plane 
\be
(\hat\sfx^1+i\hat\sfx^2)_\I = a \,\exp[2\pi i \I/\nfive]
~~,~~~~
\hat\sfx^3_\I=\hat\sfx^4_\I=0  ~~,
\ee
leads to a modified background which is also described by an exact worldsheet CFT
\be
\label{cosetorb}
\bigg( \frac{\sltwo_\nfive}{\uone}\times\frac{\sutwo_\nfive}{\uone}\bigg)\Big/ \bZ_\nfive  ~.
\ee
In this background, the radial linear dilaton is ``capped'', with a maximal value of the string coupling set by the scale $a$ of the brane separation; there is a well-defined perturbative expansion of string amplitudes.

Another way to see that string dynamics is perturbatively well-behaved is to note the worldsheet dualities
\be
\frac{\sltwo_\nfive}{\uone} \Longleftrightarrow \textrm{ N=2~Liouville}
~~~,~~~~
\frac{\sutwo_\nfive}{\uone} \Longleftrightarrow  \bZ_\nfive~\textrm{Landau-Ginzburg}
\ee
which result in an equivalent description in terms of an $N=2$ Liouville superfield $\sfX$ and Landau-Ginsburg superfield $\sfZ$, with a superpotential
\be
\label{spotl}
\cW = \sfZ^\nfive - \mu e^{-\nfive \sfX} = \prod_{\\I=1}^\nfive \Big(\sfZ - \lambda_\I e^{-\sfX} \Big)
\quad,\qquad
\lambda_\I = \mu^{\frac1\nfive}e^{\frac{2\pi i \I}\nfive}  ~.
\ee
The zeroes of the superpotential code the locations of the NS5-branes.
The Liouville wall is as usual an exponential barrier that prevents strings from penetrating into strong coupling, beyond the scale set by $\mu$.  In the geometric worldsheet dual, $\rho_{\rm min}\approx \log\mu\approx\log a$ is the scale of separation of the NS5-branes.

The non-renormalization of the superpotential~\eqref{spotl} tells us that the $\lambda_j$ are moduli of the worldsheet theory, whose deformations move the NS5's in their transverse space (in the $1$-$2$ plane).  If we allow $\lambda_i\to\lambda_j$ for some pair $i,j$ of fivebranes, then the superpotential degenerates along $\cW\sim\nabla\cW\sim0$, and a flat direction opens up; the wall keeping strings from exploring strong coupling recedes, and once again string perturbation theory breaks down (see for instance the discussion in~\rcite{Martinec:2020gkv}).  
Locally, we recover the CHS solution~\eqref{CHS geom}, which exists for any $\nfive\ge 2$.
The conclusion is that two or more coincident NS5's have a throat that is large enough for perturbative strings to fit into, while there is no such throat for individual, isolated NS5's that strings can penetrate to any significant depth.%
\footnote{The single-NS5 worldsheet theory of~\rcite{Eberhardt:2018ouy,Eberhardt:2019ywk} is not a counterexample to this notion~-- in calculations of perturbative amplitudes~\rcite{Dei:2020zui}, the string worldsheets are pinned near the $AdS_3$ boundary.  There is only a single $\sltwo$ representation in the string spectrum (and its spectral flows), and so no way to make a localized wavepacket in the $AdS_3$ bulk.}
We also see that the depth of the throat is governed by the brane separation.

\begin{figure}
\centering
\includegraphics[scale=0.5]{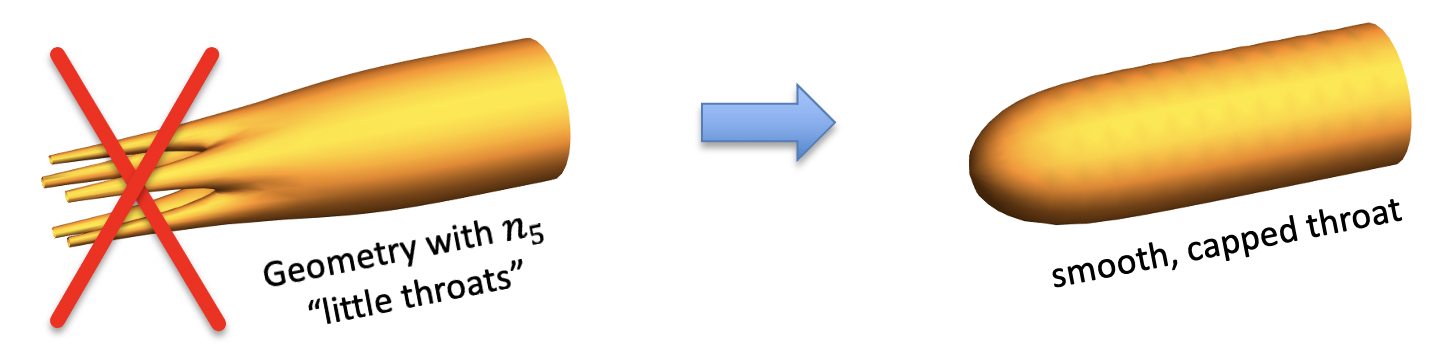}
\caption{\it A naive Coulomb branch geometry would resolve the decoupled throat of $\nfive$ NS5-branes into $\nfive$ ``little throats" at the scale set by the NS5 separation.  Strings do not fit into these purported single-NS5 throats, leading to an effective geometry capped at the scale set by the brane separation. }
\label{littlethroats}
\end{figure}

Supergravity suggests a naive picture of the circular array of separated NS5-branes which replaces the decoupling limit of the harmonic function $\sfH_5$ of~\eqref{NS5stack} by its Coulomb branch counterpart~\eqref{H5coul}.  The resulting geometry has a throat of size $\sqrt{\nfive}\,\ell_\str$ until one gets to the scale $|\hat\sfx|\sim a$, at which point the throat resolves into $\nfive$ individual throats of string scale size as depicted in figure~\ref{littlethroats}. 
%
%
But the exact solution indicates that there is no perturbative string phase space associated to such string-scale ``little throats'', so even if they could be said to exist, they would be entirely rigid, with low-energy strings unable to penetrate them, and thus they would not support local excitations.%
\footnote{In principle, D-branes might penetrate such throats, since they resolve geometry down to the Planck scale rather than the string scale; however, open strings on such probe branes are still unable to penetrate any putative throat, and so the D-brane dynamics down such a throat would again be frozen.}

The upshot is that string perturbation theory in NS5 backgrounds is well-controlled when the NS5's are well-separated.  The separation controls the depth of the NS5 throat, and the effective coupling at the geometrical, stringy cap seen by perturbative strings.

\subsection{NS5-P Lunin-Mathur geometries}
\label{sec:NS5-P intro}

Now let us consider 1/2-BPS NS5-P geometries in which one only excites waves of the scalars $\vec\sfF$ describing the transverse position of the fivebranes moving along one particular spatial direction of the fivebrane, which we parametrize by $\ytil$.  We will also compactify the fivebrane worldvolume on $\bS^1_\ytil\times \bT^4$, with $R_\ytil$ the radius of the $\ytil$-circle.
We also denote $v=t+\ytil$, $u=t-\ytil$.
We assume the geometry preserves supersymmetries assocated to Killing spinors whose product is the Killing vector $\partial_u$, so that $u$ is a Killing coordinate, and we can take all the fields to be $u$-independent.

We take the fivebrane decoupling limit~\eqref{decoupling2}, and drop the hats on the transverse space coordinates to reduce clutter.
The supergravity solution then takes the form
\begin{align}
\label{NS5-P geom}
 ds^2 &= -du\,dv + \sfH_p\, dv^2 +2\,\sfA_i\,dx^i\,dv + \sfH_5\, d\sfx^2 + ds^2_{\bT^4}
\nn\\[.3cm]
dB_2 &=  *_\perp d\sfA  \wedge dv+ *_\perp d\sfH_5  
\\[.3cm]
e^{2\Phi} &= \sfH_5 ~.
\nn
\end{align}

The geometry~\eqref{NS5-P geom} is specified by a set of harmonic forms/functions $\sfH_5,\sfA,\sfH_p$ which are given by
\begin{align}
\label{NS5-P harmfns}
\sfH_5(v,\sfx) = \sum_{\I=1}^\nfive \frac{\alpha'}{|\sfx-\sfF_\I(v)|^2}
~~,~~~
\sfA(v,\sfx) = -\sum_{\I=1}^\nfive \frac{\alpha'\partial_v\sfF_\I}{|\sfx-\sfF_\I(v)|^2}
~~,~~~
\sfH_p (v,\sfx) = \sum_{\I=1}^\nfive \frac{\alpha'(\partial_v\sfF_\I)^2}{|\sfx-\sfF_\I(v)|^2}  ~.
\end{align}
The harmonic function $\sfH_5$ is sourced by a set of $\nfive$ wiggly fivebranes located at $\sfx=\sfF_\I(v)$ with the twisted boundary condition
\be
\sfF_\I(v+2\pi R_y) = \sfF_{\I+1}(v)
~~,~~~~
\I+\nfive \sim \I 
\ee
so that all the fivebranes are bound together; in other words, we have a single fivebrane that wraps $\nfive$ times around $\bS^1_{\ytil}$.
When the notation is unambiguous, we sometimes will write the source as such a single fivebrane, parametrized by the $\nfive$-fold cover of the $\ytil$ circle.
Similarly, $\sfH_p$ is sourced by the momentum density along the wiggly fivebrane profile.

We see that the fields diverge quadratically at the fivebrane source, but our understanding of the backgrounds with exactly solvable worldsheet constructions strongly suggests that this is an artifact of the supergravity approximation, so long as the $\nfive$ windings of the fivebrane are sufficiently separated in the transverse space.  This separation will be governed by the amount of momentum excitation; roughly, the root mean square separation of the profile plays the same role as the Coulomb vev $a$ in~\eqref{H5coul}.

\begin{figure}
\centering
\includegraphics[scale=0.5]{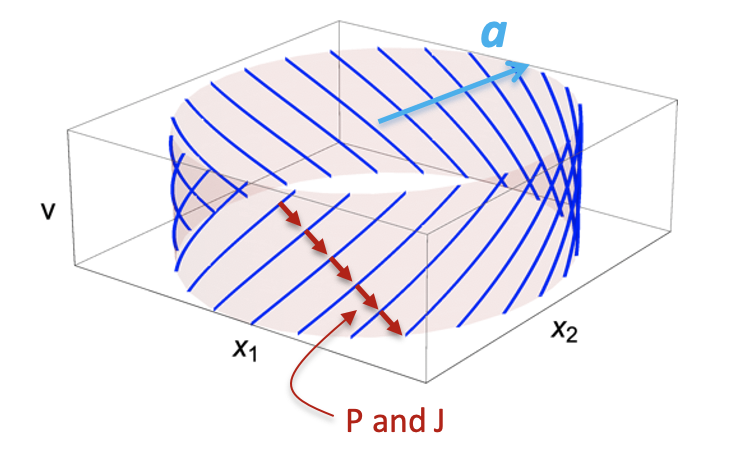}
\caption{\it Fivebranes coherently spinning in a transverse plane.  The BPS condition requires that the net momentum is transverse to the fivebrane worldvolume; the fivebranes thus carry both momentum along $\ytil$ and angular momentum in the $x^1$-$x^2$ plane, which dynamically determines the radius $a$. }
\label{fig:Circular NS5-P}
\end{figure}

An exact worldsheet description exemplifies this nonsingular behavior~\rcite{Martinec:2017ztd}; it describes strings propagating in the background sourced by the special profile
\be
\label{Fround}
\sfF^1\!+\!i\sfF^2 = a_k\,\exp\Big[{\frac{ik\sfv}{\nfive R_{\tilde{y}}}}\Big]
~~,~~~~
\sfF^3=\sfF^4=0
\ee
which describes a multiply-wound fivebrane coherently spun up in a transverse plane.  Viewed at fixed $\sfv$, one sees a ring of $\nfive$ fivebranes in a $\bZ_\nfive$ symmetric array along a circle of radius $a_k$; sitting at fixed $|\sfx^1\!+\!i\sfx^2|=a_k$ and by moving along $\bS^1_\ytil$, one crosses the fivebrane source $k$ times; see figure~\ref{fig:Circular NS5-P} for the example $\nfive=25,k=3$.  
%
%

The worldsheet CFT involves a gauged WZW model which modifies~\eqref{cosetorb} to include the longitudinal directions $t,\tilde y$ of the fivebranes
\be
\label{nullGWZW}
\frac\cG\cH = 
\frac{\sltwo_\nfive\times\sutwo_\nfive\times \bR_t\times \bS^1_\ytil}{\uone_L\times\uone_R} ~.
\ee
The parameter $k$ which determines the pitch of the fivebrane spiral specifies a discrete choice of the embedding of $\cH$ into $\cG$ which tilts the gauge orbits into the $t,\ytil$ directions.%
\footnote{The choice $k=0$ provides an alternate and equivalent construction of the circular array of Coulomb branch fivebranes~\eqref{cosetorb}.}

Thus one has essentially the same sort of Coulomb branch configuration as in~\eqref{cosetorb}, 
but now adiabatically evolving along $\sfv$, with the radius $a_k $ dynamically determined by the momentum/angular momentum carried by the fivebranes.  In particular, one has an amplitude $a_k$ for the $k$-wound source that is $\sqrt{k}$ times smaller than the amplitude $a_1$ for the $k=1$ solution, and so string dynamics is more strongly coupled at the cap for larger $k$.

In coordinates
\be
\label{bipolar}
\sfx^1+i\sfx^2 = a \cosh\rho\,\sin\theta\,e^{i\phi}
~~,~~~~
\sfx^3+i\sfx^4 = a \sinh\rho\,\cos\theta\,e^{i\psi} ~,
\ee
the geometry sourced by the profile~\eqref{Fround} is
\begin{align}
\label{round NS5-P}
ds^2 &= \Bigl( -d\sfu\, d\sfv + ds_{\scriptscriptstyle\mathbf T^4}^2 \Bigr)
+ \nfive\Bigl[d\rho^2+d\theta^2 +  \frac{1}{\Sigma} \Bigl( {\cosh}^2\!\rho\sin^2\!\theta \,d\phi^2 + {\sinh}^2\!\rho\cos^2\!\theta \,d\psi^2 \Bigr)\Bigr]
\nn\\[.1cm]
& \hskip 2cm 
+\frac{1}{\Sigma} \Bigl[ \frac{2 k }{  R_\ytil} \sin^2\!\theta \,d\sfv\, d\phi + \frac{k^2}{\nfive R_\ytil^2} d\sfv^2 \Bigr] 
\\[8pt]
B_2  
&= \frac{n_5 \cos^2\theta \cosh^2\rho}{\Sigma} d\phi \wedge d\psi  +
\frac{ k  \cos^2\theta}{ R_\ytil\,\Sigma}  d\sfv \wedge d\psi 
\quad,\qquad~~~
e^{2\Phi}  = \frac{k^2 V_4}{n_p\Sigma} ~,
\nn
\end{align}
where the denominator of the harmonic functions is
\be
\label{Sigmadef}
\Sigma \;=\; {\sinh^2\!\rho + \cos^2\!\theta}  ~.
\ee
The fivebrane ``singularity'' is spread out along the circle $\rho\!=\!0$, $\theta\!=\!\pi/2$.
Once again, worldsheet string theory is perfectly sensible, even though the supergravity fields appear to diverge, because perturbative strings cannot get close enough to the fivebrane sources to see strong coupling effects, large curvatures, \etc.

The exact solutions support the idea that the NS5-P solutions are well-behaved in perturbative string theory, whenever the fivebranes are sufficiently separated in their transverse space.

\subsection{1/2-BPS Fivebrane stars}

In~\rcite{Martinec:2023xvf,Martinec:2023gte}, the properties of generic 1/2-BPS NS5-P configurations were considered, and these states were shown to be well-characterized as ``fivebrane stars''.  The fivebranes compactified on $\bS^1_\ytil\times\bT^4$ are given a twisted boundary condition on $\bS^1_\ytil$ that combines the fivebranes into a single fivebrane wrapping that circle $n_5$ times; the scalars carry a total $n_p$ units of momentum along $\bS^1_\ytil$.  Excitations of the scalars that parametrize the fivebranes' location in the transverse space keep them apart; this separation enables the system to be self-consistently weakly coupled via the mechanism discussed above.
These 1/2-BPS NS5-P configurations are known as {\it supertubes}; the supergravity solutions that they source were elaborated in~\rcite{Lunin:2001fv,Lunin:2002iz,Kanitscheider:2007wq}.%
\footnote{We count the number of supersymmetries relative to the NS5-brane vacuum.} 

\begin{figure}[ht]
\centering
\includegraphics[scale=0.4]{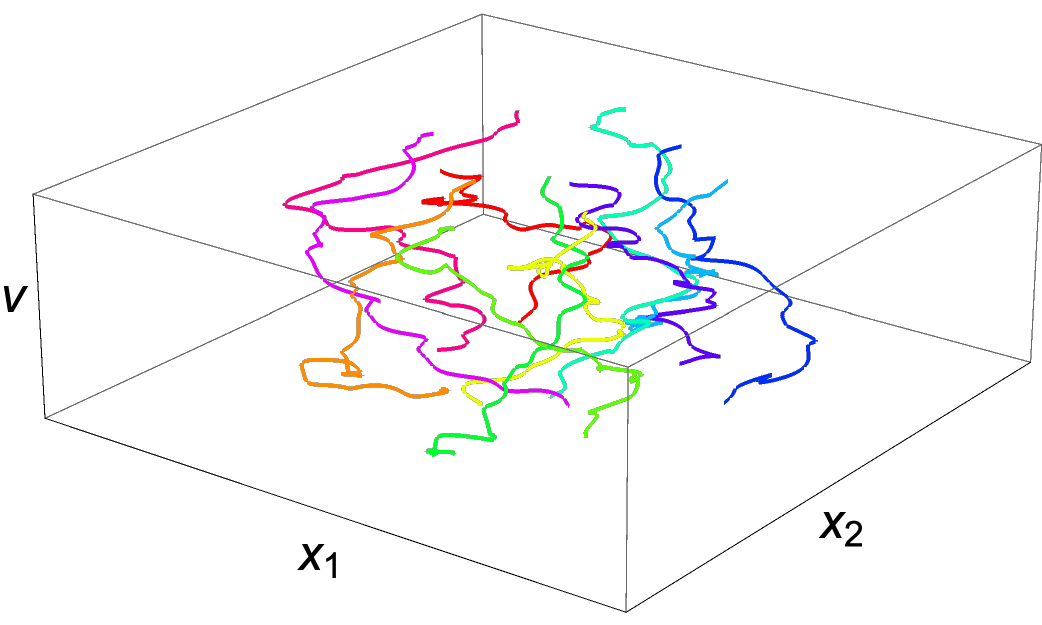}
\caption{ \it
A highly excited fivebrane multiply wrapping the vertical $\ytil$-circle executes a random walk in its transverse space.  In this depiction of a typical state, the hues of wrapped fivebrane strands evolve along the color wheel to indicate the brane connectivity.  
 }
\label{fig:wigglyNS5}
\end{figure}

Reducing along the $\bT^4$; the fivebrane becomes a magnetically charged effective string, still winding $\nfive$ times around $\bS^1_\ytil$.  This highly excited effective string executes a random walk in its transverse space, see figure~\ref{fig:wigglyNS5}.  
The coherence length of the random walk is the typical wavelength of the excitations, and the number of coherence lengths is the number of steps in the random walk.  With a total excitation level $N_\perp=n_5n_p$, the number of steps is $\sqrt{N_\perp}$ and thus the typical fivebrane star occupies a spherical blob of radius
\be
r_b = \mu \left(\frac{\pi^2 N_\perp}{6}\right)^{1/4}
~,
\ee
where
\be
\label{mudef}
\mu^2 = \frac{\alpha'^3}{V_{\bT^4}}
\ee
is the inverse tension of the effective string obtained by dimensional reduction of a fivebrane on~$\bT^4$.%
\footnote{Before taking the fivebrane decoupling limit, the scale $\mu$ is given by $\mu^2=g_s^2\frac{(\alpha')^3}{V_{\bT^4}}$.  In the process of taking the limit~\eqref{decoupling2}, we write $r_b^2/\mu^2=\hat r_b^2/\hat\mu^2$;
the factors of $g_s$ are absorbed into the rescaled coordinates and inverse tension, so that~\eqref{mudef} sets the scale in the decoupled theory (and afterward we again drop the hats to avoid clutter).}

\begin{figure}[ht]
\centering
\includegraphics[scale=0.4]{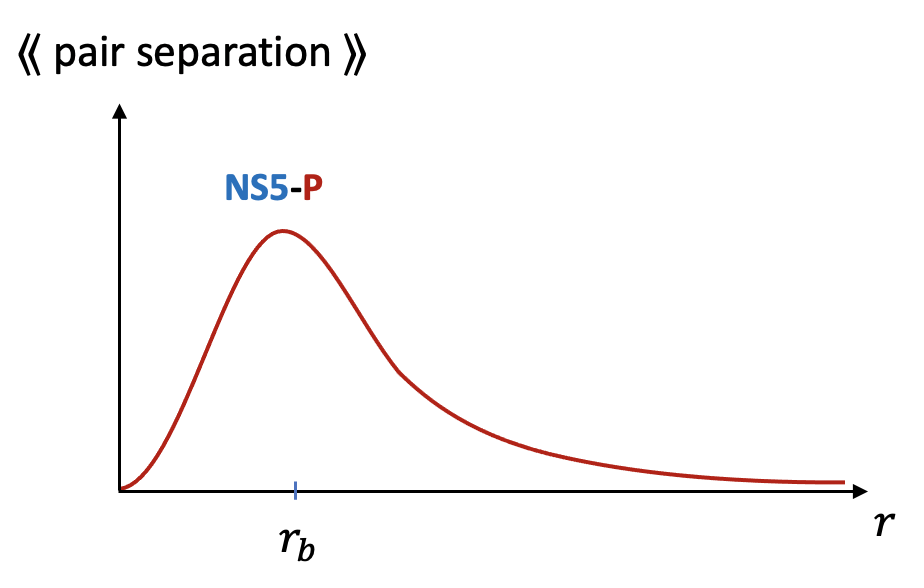}
\caption{ \it
The likelihood that points along the profile are separated by $d$ is small when $d$ is small.  The likelihood peaks at $d\approx r_b$, and decreases exponentially to zero at large scales.  
 }
\label{fig:separation}
\end{figure}

The typical element of the ensemble of such fivebrane stars is self-consistently weakly coupled.  The probability distribution for the fivebranes' separation peaks at a scale of order $r_b$ and falls smoothly to zero with decreasing separation, see figure~\ref{fig:separation}, and so the fivebrane star is self-consistently supported at weak coupling on the fivebranes' Coulomb branch.

Because the typical fivebrane star is weakly coupled, and effectively abelianized by its self-separation, it is appropriate to consider the fivebrane worldvolume theory as a source in the supergravity equations of motion.  We carry out a general analysis of 1/2-BPS NS5-P field configurations in section~\ref{sec:STsolns}, varying the supergravity action coupled to the NS5-brane worldvolume source action, and imposing the general form of 1/2-BPS solutions.  These solutions were originally derived in~\rcite{Lunin:2001fv,Lunin:2002iz,Kanitscheider:2007wq} using U-duality transformations on the F1-P solutions of~\rcite{Callan:1995hn,Dabholkar:1995nc}; we will instead construct them here directly in the NS5-P duality frame in order to introduce the effective action formalism that lies at the heart of our analysis.

In addition to the transverse excitations of the fivebrane, there are 1/2-BPS excitations of the gauge field multiplet on the fivebrane.  D-branes can end on NS5-branes, and are charged sources for the NS5 gauge multiplet; thus these gauge fields have odd R-R parity, and their excitations on the fivebrane also source R-R fields in the bulk.

Because the fivebrane strands are well-separated, at low energies we can restrict our consideration to the abelian subsector of the full non-abelian fivebrane gauge dynamics. 
The NS5-P solution~-- including such abelian internal gauge excitations, and the bulk R-R fields they source~-- is derived in section~\ref{sec:gauge profiles}; the explicit solution is written in Eqs.~(\ref{gen STube NS})-(\ref{HpRR}).  While the solutions have singularities, and regions of large field gradients, everything we understand about the worldsheet theory indicates that these are artifacts of the supergravity approximation, and perturbative string theory is perfectly self-consistent.

Aspects of typical 1/2-BPS configurations are captured by the ensemble average over all such geometries.  The ensemble average over transverse fivebrane excitations was analyzed in~\rcite{Alday:2006nd,Balasubramanian:2008da,Raju:2018xue,Martinec:2023xvf}; we review the results of this analysis and extend the discussion to include internal fivebrane excitations in section~\ref{sec:NS5-P ensemble}.  In particular, in section~\ref{sec:density 2pt} we evaluate the two-point correlator of the fivebrane charge density and find large statistical fluctuations, which are the source of the large fluctuations in the metric coefficients observed in~\rcite{Raju:2018xue}; and in section~\ref{sec:graviton 2pt}, we compute the contribution to the graviton two-point correlator mediated by the fivebrane source dynamics.

\paragraph{Near-BPS moduli space dynamics:}
When the system is excited away from the BPS bound, the geometry is more complicated, since it is no longer severely constrained by the form of 1/2-BPS solutions.  However, one can expand the effective action around the BPS limit.  Our intuition here follows from the treatment of monopole dynamics, where the dynamics is governed by a sigma model on the moduli space of BPS multi-monopole solutions.  Here, we expect slow dynamics on the 1/2-BPS configuration space.  We should be able to use the same effective gravity+brane action, but now   keeping the leading order dependence on non-BPS quantities.

Where should we end up?  We expect the excess energy to cause the system to ergodically explore the BPS state space for low excitation levels, and for sufficiently high excitation to undergo slow gravitational collapse.  Already in the BPS configuration space, one can cause the fivebrane star to shrink by transferring the excitation budget into 1/2-BPS internal gauge field modes on the NS5’s; increasing the average mode number of transverse scalar excitations has a similar effect.  These modifications result in a more compact fivebrane profile, and since the transverse extent of the profile governs the scale of the cap in the geometry, the throat deepens and the redshift to the cap increases.  Eventually one reaches the scale at which fluctuations in the fivebrane location are no longer suppressed; the fivebranes merge and non-abelian fivebrane dynamics takes over.  This is the onset of the black hole phase. 

We discuss aspects of perturbations of 1/2-BPS fivebrane stars in section~\ref{sec:perts}, and a variety of future projects in section~\ref{sec:discussion}.

\section{Solving for NS5-P supertubes}
\label{sec:STsolns}

\subsection{The fivebrane effective action}
\label{sec:NS5 effact}

We now want to show how the geometry~\eqref{NS5-P geom} solves the equations of motion of type II supergravity coupled to NS5-brane sources.  To this end, we consider the type II supergravity effective action, and for this subsection we restrict our attention to the NS-NS sector fields $G,B^{(2)},\Phi$. The next subsection incorporates R-R fields.

The 10D supergravity action for the NS sector fields $G,B_2,\Phi$ is given by
\begin{equation}
\label{NSact}
\cI_{\rm\sst NS\cdot NS} = \frac{1}{2\kappa_0^2}\int\! d^{10} x\, e^{-2\Phi} \sqrt{-G} \left(R+4\nabla_{\mu} \Phi \nabla^{\mu} \Phi -\frac{1}{12} H_{\alpha \beta \gamma} H^{\alpha \beta \gamma} \right)~,		
\end{equation}
with $H_3=dB_2$ and
\begin{equation}
\label{kappadef}
\kappa_0^2 = \kappa ^2 /g_s ^2 = 2^6 \pi^7 (\alpha')^4
~~,~~~~ 
\kappa^2 = 8\pi G_N ^{\sst (10d)}~.
\end{equation}
Most of this subsection we work with a dimensional reduction of that action on $\bT^4$, taking the target space fields to be independent of the torus directions. The 6d action is then
\begin{equation}
\label{NSact2}
\cI_{\rm\sst NS\cdot NS} = \frac{V_{\bT^4}}{2\kappa_0^2}\int\! d^{6} x\, e^{-2\Phi} \sqrt{-G} \left(R+4\nabla_{\mu} \Phi \nabla^{\mu} \Phi -\frac{1}{12} H_{\alpha \beta \gamma} H^{\alpha \beta \gamma} \right)~.
\end{equation}
NS5-branes are magnetically charged under the NS two-form potential ${B_2}$.  The fivebrane worldvolume action couples directly to the dual 6-form potential $\widetilde B_6$
\begin{equation}
\label{Kdef}
K = e^{-2\Phi} *H ~~,~~~~ K = d{\widetilde B} ~,
\end{equation} 
so we will also be interested in the action written in terms of $\widetilde{B}_6$
(see e.g. \rcite{Duff:1994vv})
\begin{equation}
\cI_{\rm\sst NS\cdot NS} = \frac{V_{\bT^4}}{2\kappa_0^2}\int\! d^{6} x\, e^{2\Phi}\sqrt{-G}\Big(  R+4\nabla_{\mu} \Phi \nabla^{\mu} \Phi -\frac{1}{2\times 7!} K_{\alpha \beta \gamma \delta \epsilon \zeta \eta} K^{\alpha \beta \gamma \delta \epsilon \zeta \eta} \Big)~.		
\end{equation}
%
%
The NS5-brane has a ``Nambu-Goto'' style effective kinetic term together with a coupling to a six-form potential $\widetilde{B}_6$ which is the magnetic dual of $B_2$~\rcite{Eyras:1998hn}
\begin{equation}
\label{NS5act no-RR}
\cS_{\rm source} = -T_{A5} \int\! d^6 \xi\,  \left( e^{-2\Phi} \sqrt{-G } +
\frac{1}{6!}\widetilde{B}_6
\right)~,
\end{equation}
where $T_{A5}$ is the tension of a IIA fivebrane.  
We can perform a dimensional reduction on $\bT^4$;
as long as one is suppressing winding and momentum modes along the torus directions (which should be absent in the Wilsonian effective action with a string scale $\bT^4$), nothing in the dynamics depends on $\bT^4$, and so such a dimensional reduction is appropriate.
Under this dimensional reduction, the fivebrane becomes an effective string in 6d, coupling to the effective two-form potential $\widetilde B_2$ which is the 6-form potential $\widetilde B_6$ with four of its legs along $\bT^4$ (which we suppress in what follows).

The fivebrane effective action becomes an effective string action 
\begin{equation}
\label{Reduction}
\cS_{\rm source} = 	-\tau_{\rm\sst NS5}\int d^2\sigma  \left(  e^{-2\Phi}\sqrt{-G  }+\frac{1}{2}\epsilon^{ab}\widetilde{B} _{ ab} \right)~
\end{equation}
in a 6d target spacetime,
where the tension is given by \rcite{Polchinski2}
\begin{equation}
\tau_{\rm\sst NS5} = \frac{2\pi^2 \alpha' V_{\bT^4}}{\kappa_0^2} ~.
\end{equation}
We can write this effective action in a Polyakov-like form by introducing an intrinsic worldvolume metric~$\gamma_{ab}$%
\footnote{ The first term on the R.H.S is similar to Eq.~(6.1) of reference \cite{Callan:1991ky} for which the intrinsic metric is flat.  Reference \rcite{Duff:1994vv} also arrived at the same action Eq.~(\ref{Source}).}
\begin{align}
\label{Source}
	\cS_{\rm source} &=-\frac{1}{2}\tau_{\rm\sst NS5} \int d^{2} \sigma \sqrt{-\gamma_2}  \left(\gamma^{ab} e^{-2\Phi}G_{\mu \nu}\partial_a \sfF^{\mu} \partial_b \sfF^{\nu}+\epsilon ^{ab} \widetilde{B}_{\mu \nu}\partial_a \sfF^{\mu} \partial_b \sfF^{\nu}\right)~.
\end{align}
Varying this action with respect to $\gamma_{ab}$ results in the reparametrization constraints 
\begin{equation}
\label{Constraints}
G_{\mu \nu} \partial_{a} \sfF^{\mu} \partial_{b} \sfF ^{\nu} = \frac{1}{2}G_{\mu \nu} \gamma_{ab} \gamma^{cd} \partial_c \sfF^{\mu} \partial_d \sfF^{\nu}~. 
\end{equation}
Eq.~(\ref{Constraints}) implies that the intrinsic metric is proportional to the pullback metric, and plugging it into Eq.~(\ref{Source}) brings us back to Eq.~(\ref{Reduction}).

We will work in the conformal gauge $\gamma_{ab}=\eta_{ab}$ on the effective string worldsheet.  The equation of motion obtained by varying the action with respect to $\sfF^{u}$ is then $\partial_+\partial_- \sfF^v=0$.  BPS solutions will have $\sfF^v=\sfF^v(\sigma_+)$, and by a conformal transformation we set $\sfF^v=\sigma_+=v$.
Then for $a=b=+=v$, the constraint equation reads
\begin{equation}
\label{vvConstraint}
\sum_{\sfm=1} ^{n_5} \left( -\partial_v \sfF^u _m +H_5 |\partial_v \vec{\sfF}_\sfm (v)|^2+2\vec{\sfA} \cdot \partial_v \vec{\sfF}_\sfm (v) + H_P   \right) = 0~. 
\end{equation}
Note that with the harmonic functions~\eqref{NS5-P harmfns}, this expression is naively singular due to the self-interaction of the fivebranes; however, the fivebrane self-energy should cancel between bosonic and fermionic contributions due to supersymmetry (see~\rcite{Michel:2014lva} for a related discussion).  Henceforth we will ignore such singular terms in the effective action.

Varying the action~\eqref{Source} with respect to the transverse embedding fields $\sfF^i$, one finds the equations of motion
\begin{align}
\nabla^2 \sfF^{i} &=\Big(-\tilde{\Gamma}^i _{\alpha \beta} \gamma^{ab} +\frac{1}{2} \epsilon^{ab} e^{2\Phi} K^{i} _{~~\alpha \beta} \Big) \partial_a \sfF^{\alpha} \partial_b\sfF^{\beta} ~,
\end{align}
where $\nabla^2$ is the Laplacian on the worldsheet, $\widetilde{\Gamma}$ denotes the Chritoffel symbol of the metric $e^{-2\Phi} G_{\mu \nu}$, and $K=d \widetilde{B}$. In the conformal gauge and the BPS ansatz~\eqref{NS5-P geom} for the target space fields, one finds
\begin{align}
-2\partial_+ \partial_- \sfF^i &=\Big(2\tilde{\Gamma}^i _{\alpha +}  + \epsilon^{-+} e^{2\Phi} K^{i} _{~~\alpha +} \Big) \partial_- \sfF^{\alpha}  ~,
\end{align}
and a calculation of the R.H.S. implies that it vanishes due to cancellations between terms that involve $\partial_u \sfF^i $, and terms proportional to $\partial_i \frac{1}{H_5}$. Therefore, any $\sfF^i(v)$ is a solution.

Finally, the equation of motion obtained from the variation of $\sfF^v$ then reads
\be
-2\partial_+\partial_- \sfF^u = \Big(2\tilde{\Gamma}^u _{\alpha +}  + \epsilon^{-+} e^{2\Phi} K^{u} _{~~\alpha +} \Big) \partial_- \sfF^{\alpha}  ~.
\ee
Note that 
\be
2\tilde{\Gamma}^u _{uv} + \epsilon^{uv}e^{2\Phi} g^{ui} \partial_i \widetilde{B}_{uv}=0 ~,
\ee
therefore the coefficient of $\partial_-\sfF^u$ on the RHS vanishes.
Then since $\partial_-\sfF^v\tight=\partial_- \sfF^i\tight=0$, $\sfF^u$ is also a free field.
We saw above that its left-moving part was determined by the constraint~\eqref{vvConstraint}.  We can then use a reparametrization to set the right-moving part to $\partial_- \sfF^u(\sigma_-)=1$.

Introducing a unit integral $\int\! d^4 \sfx\, \delta^{4}(\sfx - \sfF(v))=1$, results in a 6D effective string action 
\begin{align}
\label{Source2}
	\cS_{\rm source} &=-\frac{1}{2}\tau_{\rm\sst NS5} \int d^{6} x   \left(\sqrt{-\gamma_2}\,\gamma^{ab} e^{-2\Phi}G_{\mu \nu}\partial_a \sfF^{\mu} \partial_b \sfF^{\nu}+\epsilon ^{ab} \widetilde{B}_{\mu \nu}\partial_a \sfF^{\mu} \partial_b \sfF^{\nu}\right)\delta^{4}(\sfx - \sfF(v)) ~.
\end{align}
This is the action for a single fivebrane; recall that we have $\nfive$ source strands cyclically wound together.


The 1/2-BPS condition allows us to restrict the form of the solution to~\eqref{NS5-P geom}.  To verify the solution~\eqref{NS5-P harmfns}, we vary the bulk and brane actions~\eqref{NSact2}, \eqref{Source2} with respect to the spacetime fields $G,B,\Phi$ and solve the resulting equations of motion.  Note, however, that the metric ansatz admits a residual diffeomorphism invariance under which
\be
u\to u' = u +2 \Lambda(v,\sfx)
~~,~~~~
\sfA\to \sfA' = \sfA + \nabla\Lambda
~~,~~~~
\sfH_p \to \sfH_p' = \sfH_p + 2\partial_v\Lambda  ~.
\ee
We use this residual symmetry to fix the gauge
\be
\label{Agauge}
\partial_v H_5 = \nabla\cdot\sfA ~,
\ee
which simplifies some of the equations of motion.

The resulting equations of motion are as follows:  The dilaton equation is given by
\begin{align}
\label{dilaton2}
		R - 4\nabla_{\alpha} \Phi \nabla^{\alpha} \Phi +4\nabla^2 \Phi-\frac{1}{12} H_{\alpha \beta \gamma } H^{\alpha \beta \gamma }&= \frac{1}{\sqrt{-G}} \kappa_0^2 \tau_{\NSfive} \sqrt{-\gamma_2}  \gamma^{ab} G_{\mu \nu}\partial_a \sfF^{\mu} \partial_b \sfF^{\nu}\delta^{4}(\sfx - \sfF(v))~.
\end{align}
The electric Kalb-Ramond equation of motion
\begin{equation}
\label{ElectricBeom}
	\partial_{\mu} \left( \sqrt{-G}e^{-2\Phi} H^{\mu \alpha \beta}\right)=0
\end{equation}
is the Bianchi identity for the dual magnetic Kalb-Ramond equation
\begin{equation}
	\label{MagneticBeom}
	\partial_{\mu} \left(e^{2\Phi} \sqrt{-G}    K^{\mu \beta \gamma } \right)=2\kappa_0 ^2 \tau_{\NSfive}\, \epsilon^{ab} \partial_a \sfF^{\beta} \partial_b \sfF^{\gamma}  \delta^4 (\sfx- \sfF(v)) ~,
\end{equation}
where $K$ was defined in Eq.~\eqref{Kdef}.
The metric equations of motion are as usual
\be
\mathcal{G}_{\mu \nu} = \kappa_0 ^2 e^{2\Phi} T_{\mu \nu}
\ee
where
\begin{equation}
\label{comb2}
		\mathcal{G}_{\mu \nu}\equiv R_{\mu \nu} + 2\nabla_{\mu}  \nabla_{\nu} \Phi -\frac{1}{4}  H_{\mu \alpha \beta } H_{\nu } ^{~~\alpha \beta }-\frac{1}{2} G_{\mu \nu} \left( R - 4\nabla_{\alpha} \Phi \nabla^{\alpha} \Phi +4\nabla^2 \Phi-\frac{1}{12} H_{\alpha \beta \gamma } H^{\alpha \beta \gamma }\right)
		~.
\end{equation}
We will calculate below the components of $T_{\mu \nu}$ by varying the fivebrane source action with respect to the metric.

Let us now systematically solve these equations of motion.
The dilaton equation (\ref{dilaton2}) for the ansatz of Eq.~(\ref{NS5-P geom}) reduces to
\begin{equation}
\label{dilatonresult}
	\nabla^2 \sfH_5 (\sfx,v)=-4\pi^2 \alpha' \sum_{\I=1} ^{n_5} \delta^4 (\sfx-\sfF _{\I}(v)) ~, 
\end{equation} 
whose solution is
\begin{equation}
\label{H5+const}
\sfH_5 (\sfx,v) = 1+\sum_{\I=1} ^{n_5} \frac{\alpha'}{|\sfx-\sfF_{\I}(v)|^2}~.
\end{equation}
The constant term results from the imposition of asymptotically flat boundary conditions.  We are interested in the fivebrane decoupling limit analogous to~\eqref{decoupling}.  Therefore, from now on we take the NS5-brane decoupling limit~\eqref{decoupling}: $g_s \to0$, $\vec{\,\sfx}\to0$ with $\frac{\!\vec{\,\sfx}}{g_s}$ and $\frac{\!\vec{\,\sfF}}{g_s}$ fixed. In this limit, the constant scales out of the fivebrane harmonic function 
\begin{equation}
\label{H5result}
  \sfH_{5} (\sfx,v) = \sum_{\I=1} ^{n_5} \frac{\alpha'}{|\sfx-\sfF_{\I}(v)|^2}~,
\end{equation}
and all explicit factors of $g_s$ disappear from the supergravity solution.
The $B$-field~(\ref{ElectricBeom}) is automatically satisfied for the configuration in~(\ref{NS5-P geom}), because there are no electric sources.  The equations of motion obtained by varying the magnetic potential $\widetilde B$ are either trivially satisfied by the ansatz~\eqref{NS5-P geom} or (for the $u$-$v$ component) equivalent to the dilaton equation~\eqref{dilatonresult}.

Next, we move to the metric equations for (\ref{NS5-P geom}) and compute the LHS which we called $\mathcal{G}_{\mu \nu}$.  The relevant equations are
\begin{align}
\label{EinsteinEqs}
\mathcal{G}_{vi} = -\frac{\sfH_5\nabla^2 \sfA_i -\sfA_i \nabla^2 \sfH_5}{2\sfH_5^2}
~~&,~~~~
	\mathcal{G}_{vv} = -\frac{\sfH_5 \nabla^2 \sfH_p-\sfH_p \nabla^2 \sfH_5}{2\sfH_5 ^{\,2}}~,
\end{align}
where we have used the gauge~\eqref{Agauge}.
Now, the stress tensor is defined by
\begin{equation}
	T^{\mu \nu} = \frac{2}{\sqrt{-G}} \frac{\delta \cS_{\rm multi}}{\delta G_{\mu \nu}}=-\frac{\tau_{\NSfive} }{\sqrt{-G}}e^{-2\Phi}   \sqrt{-\gamma}\gamma^{ab} \sum_{\I=1} ^{n_5}\partial_a \sfF_\I^{\mu} \partial_b \sfF_\I^{\nu} \delta^4 \left(\sfx-\sfF_{\I}(v)\right)~.
\end{equation}
The worldsheet metric is the the target-space metric in the $u$-$v$ directions, up to an overall multiplicative factor. 
The components $T^{\mu \nu}$ that will give rise to new non-trivial equations
\begin{align}
\label{Tmunu}
    T^{uu} &= -2\frac{\tau_{\NSfive}}{\sqrt{-G}} e^{-2\Phi} \sqrt{-\gamma}\gamma^{uv} \sum_{\I=1} ^{n_5} \partial_v \sfF^u _m \delta^4 (\sfx-\sfF_{\I}(v))
\nn\\
\hskip 2cm
T^{uv} &= -\frac{\tau_{\NSfive}}{\sqrt{-G}} e^{-2\Phi} \sqrt{-\gamma}\gamma^{uv} \sum_{\I=1} ^{n_5} \delta^4 (\sfx-\sfF_{\I}(v))
\\
\hskip 2cm
	T^{ui} &= -\frac{\tau_{\NSfive}}{\sqrt{-G}}e^{-2\Phi}\sqrt{-\gamma}\gamma^{uv}\sum_{\I=1} ^{n_5}\partial_v \sfF_\I ^{i} \delta^4 (\sfx-\sfF_{\I}(v))~.
\nn    
\end{align}
The transverse $i$-$j$, $u$-$u$, $u$-$i$ metric component equations are trivially satisfied. 
The $u$-$v$ metric equation is equivalent to Eq.~(\ref{dilatonresult}).

We next consider the $v$-$i$ metric equations. 
The stress tensor $v$-$i$ components are
\begin{equation}
	T_{vi} = G_{vu} G_{iv} T^{uv}+G_{vu}G_{ij}T^{uj}=-\sum_{\I=1} ^{n_5}\Big(\sfA_i+\partial_v \sfF_i ^{\I} \sfH_5\Big)\frac{\tau_{\NSfive}}{\sfH_5 ^{\,2}} e^{-2\Phi} \delta^4 (\sfx-\sfF_{\I}(v))~,
\end{equation} 
and the $v$-$i$ metric equation of motion reduces to 
\begin{equation}
	\nabla^2 \sfA_i = 2\kappa_0^2\,\tau_{\NSfive}\sum_{\I=1} ^{n_5}\partial_v \sfF_i ^{\I} \delta^4 (\sfx-\sfF_{\I}(v))~.
\end{equation}
It follows that
\begin{equation}
\label{Aresult}
	\sfA(v,\sfx) = -\sum_{\I=1} ^{n_5}\frac{\alpha'\partial_v \sfF_\I }{|\sfx-\sfF_{\I}(v)|^2}~.
\end{equation}
Next, the $v$-$v$ component of the stress-energy tensor is given by
\begin{align}
\label{Tvv} 
T_{vv}&=2 G_{vu} G_{vv} T^{uv}+2G_{vi}G_{vu}T^{iu}+G_{vu}^2 T^{uu}
\\[.2cm]
&= -\sum_{\I=1} ^{n_5}\left(\sfH_p-|\partial_v \sfF(v)|^2 H_5\right)  \frac{\tau_{\NSfive}}{\sqrt{-G}} e^{-2\Phi} \delta^4 (\sfx-\sfF_{\I}(v))~.
\nn
\end{align}
Then the $v$-$v$ metric equation of motion can be written as
\begin{equation}
	\nabla^2 \sfH_p = -2\kappa_0^2 \tau_{\NSfive} \sum_{\I=1} ^{n_5}|\partial_v \sfF_{\I}|^2 \delta^4 (\sfx-\sfF_{\I}(v))~.
\end{equation}
Therefore,
\begin{equation}
	\sfH_p (v,\sfx) = \sum_{\I=1} ^{n_5}\frac{\alpha' |\partial_v \sfF_{\I} (v)|^2}{|\sfx-\sfF_{\I}(v)|^2}~.
\end{equation}

To recapitulate, we have verified that the metric, $B$-field and dilaton in Eq.~(\ref{NS5-P geom}) are solutions to the equations of motion with the harmonic functions in Eq.~(\ref{NS5-P harmfns}), sourced by the multi-center fivebrane system.  
Previously, \cite{Lunin:2001fv,Kanitscheider:2007wq} found such solutions using U-duality between wrapped fivebranes and fundamental strings.  One of our motivations for revisiting this problem is to extend the discussion to 1/4-BPS and near-BPS solutions (see sections~\ref{sec:near-BPS} and~\ref{sec:discussion} for a preliminary discussion), for which we need the effective action including sources, in the appropriate duality frame where the charges are NS5-P.

\subsection{Adding the gauge field modes}
\label{sec:gauge profiles}

We now generalize the discussion to include excitations of the gauge multiplet of the fivebranes, which source R-R fields in the bulk.
In Type IIA superstring theory, these internal modes are a scalar that we denote by $\mfa^{(0)}$ and a chiral two-form $\mfa^{(2)}$, the latter has an associated self-dual three-form field strength defined in Eq.~(\ref{curlyH}). 

We will see that the solution generalizing~\eqref{NS5-P geom}, \eqref{NS5-P harmfns} has a similar form for the NS fields,
\begin{align}
\label{gen STube NS} 
	ds^2 &= -dudv +\bigg(\sfH_p  -\frac{(\sfZ_0) ^{2} + (\sfZ_\asym )^2}{\sfH_5} \bigg)dv^2 +2\sfA_i dvdx^i+ \sfH_5 d\sfx^2+ ds_{\mathcal{M}}^2 
\nn\\[.2cm]
	B^{(2)} &= \mathbf{b} + \sfB^{(1)} \wedge dv
    ~~,~~~~
d\mathbf{b} = *_4 d\sfH_5 ~~,~~~~ d\sfB ^{(1)} = *_4 d\sfA^{(1)}
\\[.2cm]
	e^{2\Phi}&=\sfH_5 
\nn
\end{align}
together with non-trivial R-R fields
\begin{align}
\label{gen STube RR}
	C^{(1)} &= -\frac{\sfZ_0}{\sfH_5} dv
\nn\\[.2cm]
	C^{(3)} &= \biggl(\mathbf{c}+\frac{\sfZ_\asym}{\sfH_5} \Omega_\asym \biggl) \wedge dv
 \\[.2cm]
 C^{(5)} &= \frac{\sfZ_0}{\sfH_5} dv \wedge \widehat{\text{vol}}_4+  \mathbf{c}_{\asym}\wedge dv\wedge \Omega_\asym 
\nn\\[.3cm]
    d\mathbf{c} &= *_4 d\sfZ_0 ~~,~~~~ d\mathbf{c}_{\asym} = *_4 d\sfZ_{\asym}   ~, 
\nn
\end{align}
where $\Omega_\asym$ are a basis of self-dual two-forms on $\bT^4$ defined in Eq.~\eqref{2forms} below, $\widehat{\text{vol}}_4$ is the volume form on the $\bT^4$
and $\sfZ_0,\sfZ_\asym$ are R-R harmonic functions sourced by the 6d effective string profile functions $\sfF,\mfa^{(0)},\mfa^{(2)}$ as follows:
\begin{align}
\label{RRSources}
    \sfZ_0 (v,\sfx) &= -\sum_{\I=1} ^{n_5}\frac{ \alpha' \partial_v \mfa_{\I} ^{(0)}(v) }{|\sfx-\sfF_\I(v)|^2}
\nn\\[.2cm]
\sfZ_{\asym} (v,\sfx) &= -\sum_{\I=1}  ^{n_5}\frac{ \alpha' \partial_v\mfa _{\I}^{\asym} (v)}{|\sfx-\sfF_\I (v)|^2}~;
\end{align}
also, the NS field harmonic function $\sfH_p$ is modified to
\begin{equation}
\label{HpRR}
\sfH_p (v,\sfx) = \alpha'\sum_{\I=1} ^{n_5}  \frac{  \big|\partial_v \sfF_\I\big|^2 + \big|\partial_v \mfa_{\I}^{(0)}\big|^2 + \big|\partial_v \mfa_{\I}^{\asym}\big|^2 }{|\sfx-\sfF_\I(v)|^2}~.
\end{equation}
Note that the $v$-$v$ metric component in Eq.~(\ref{gen STube NS}) involves a subtraction by a quadratic form of the R-R harmonic functions. 
We will again derive the above solution by solving the equations of motion derived from the Type IIA supergravity action plus the effective action of 1/2-BPS fivebranes, generalized to include the supergravity R-R fields and worldvolume internal gauge fields.

The full IIA supergravity action is, including R-R fields, 
\begin{equation}
	\cI_{SUGRA} = \cI_{NS} + \cI _R+\cI_{CS}~,
\end{equation}
where
\begin{align}
	\cI_{NS} &= \frac{1}{2\kappa_0^2} \int d^{10} x ~ \sqrt{-G} e^{-2\Phi} \left(R+4\partial_{\mu} \Phi \partial^{\mu} \Phi -\frac{1}{12} H_{\alpha \beta \gamma} H^{\alpha \beta \gamma}\right)
\nn\\[.2cm]
	\cI_R &= -\frac{1}{4\kappa_0^2}\int d^{10} x \sqrt{-G} \left(\frac{1}{2} F_{\mu \nu} F^{\mu \nu}+\frac{1}{24} \widetilde{F}^{\alpha \beta \gamma \delta} \widetilde{F}_{\alpha \beta \gamma \delta }\right)
\\[.2cm]
	\cI_{CS} &= \frac{1}{4\kappa_0^2} \int  H^{(3)} \wedge C^{(3)} \wedge F^{(4)} ~.
\nn
\end{align}
We use a Chern-Simons term that differs from the usual Chern-Simons term $-\frac{1}{4\kappa_0 ^2} \int B^{(2)}\wedge F^{(4)} \wedge F^{(4)} $ by a boundary term, because in the presence of NS5-branes, the Kalb-Ramond field cannot be defined globally because the quantity $dH$ is nonzero and supported on the worldvolume of these magnetically-charged objects. 
We will write the relevant terms in the source action $\cS_{\rm NS5}$ of Eq.~\eqref{GeneralAction} below.
We adopt a convention for $\widetilde{F}_4$ where the second term on the R.H.S is minus that of \rcite{Polchinski2}
\begin{equation}
	\widetilde{F}^{(4)} = dC^{(3)} - H^{(3)} \wedge C^{(1)} ~. 
\end{equation} 
The dilaton equation remains unchanged (\ref{dilaton2}). The same is true for the $\widetilde{B}$ equation (\ref{MagneticBeom}).
The metric equations are
\begin{align}
e^{-2\Phi}\cG_{\mu\nu}
-\frac{1}{2} F_{\mu \alpha} F_{\nu} ^{~\alpha} -\frac{1}{12} \widetilde{F}_{\mu\alpha \beta \gamma} \widetilde{F}_{\nu} ^{~\alpha \beta \gamma}+\frac{1}{8} G_{\mu \nu} F_{\alpha \beta} F^{\alpha \beta} +\frac{1}{96} G_{\mu \nu} \widetilde{F}_{\alpha \beta \gamma \delta} \widetilde{F}^{\alpha \beta \gamma \delta}=\kappa_0^2 T_{\mu \nu} ~, 
\end{align}
where $\cG_{\mu\nu}$ was defined in~\eqref{comb2}.
Varying the R-R one-form potential leads to
\begin{align}
\label{RR1}
-\frac{1}{2\kappa_0^2 \sqrt{-G}} \partial_{\mu} \left(\sqrt{-G} F^{\mu \nu}\right)
&=-\frac{1}{12\kappa_0^2} H_{\alpha \beta \gamma} \left( F ^{\nu \alpha \beta \gamma } + C^{\nu} H^{\alpha \beta \gamma} - 3 C^{\alpha} H^{\nu \beta \gamma} \right)+\frac{\delta \cS_{\rm source}}{ \delta C_{\nu} ^{(1)}}~.
\end{align}
Varying the R-R three-form potential gives rise to
\begin{align}
\label{C3Eq}
	&-\frac{1}{12\kappa_0^2} \partial_{\mu} \left( \sqrt{-G} \widetilde{F} ^{\mu \alpha \beta \gamma }\right) + \frac{1}{(4!)^2\times 3\kappa_0 ^2} \sqrt{-G} \epsilon^{\alpha \beta \gamma \mu \lambda_1 ...\lambda_6 } \left( H_{ \lambda_1 \lambda_2 \lambda_3} (\widetilde{F}) _{\mu\lambda_4 \lambda_5 \lambda_6}+\frac{1}{2} (dH)_{\mu \lambda_1 \lambda_2 \lambda_3} C_{\lambda_4 \lambda_5 \lambda_6}\right)\nonumber\\
    &\hskip 5cm=\frac{\delta \cS_{\rm source} }{\delta C^{(3)} _{\alpha \beta \gamma}}~.
\end{align}
The variation of the action with respect to the $B$-field gives 
\begin{align}
&-\frac{1}{4\kappa_0^2} \partial_{\mu} \left( \sqrt{-G} e^{-2\Phi} H^{\mu \alpha \beta} \right) +\frac{1}{8\kappa_0^2 }\sqrt{-G}(F_2)_{\mu \nu}\widetilde{F} ^{\mu \nu \alpha \beta} 
\nn\\
&\hskip 2cm 
+\frac{1}{8\kappa_0^2 \times (4!)^2} \sqrt{-G}\epsilon^{\alpha \beta \alpha_1...\alpha_4 \alpha_5 ...\alpha_8} (\widetilde{F}_4)_{\alpha_1 ... \alpha_4} (\widetilde{F}_4) _{\alpha_5 ... \alpha_8} 
\\
&\hskip 3cm 
-\frac{1}{8\kappa_0^2 \times (4!)^2} \sqrt{-G}\epsilon^{\alpha \beta \alpha_1...\alpha_4 \alpha_5 ...\alpha_8} C_{\alpha_1 \alpha_2 \alpha_3} (dF) _{\alpha_4 ... \alpha_8}
= \frac{\delta \cS_{\rm source} }{\delta B^{(2)} _{\alpha \beta }}~.  
\nn
\end{align}
The Bianchi identities away from magnetic fivebrane sources are
\begin{equation}
dF_4=0   
~~,~~~~ 
dF_2 =0 
~~,~~~~
dH_3=0
\end{equation}
The equations of motion we derive imply that both $dF_4$ and $dH_3$ are proportional to a transverse delta function at fivebrane sources; an explicit equation for $dH_3$ in given below in Eq.~(\ref{dH}).

Most of the equations we verified in the previous subsection remain exactly the same: The dilaton, transverse metric, $u$-$\mu$ metric, $\widetilde{B}_{\mu \nu}$ and $B_{\mu \nu}$ equations.
Basically the fact the R-R fields possess a $v$-leg, coupled with the fact that $G^{vv}=0$, makes $F_{\mu \nu} F^{\mu \nu} =0,F_{\mu \nu \alpha \beta} F^{\mu \nu \alpha \beta}=0$ and $F_{\mu \nu} F^{\mu \nu \alpha \beta}=0$.
The new non-trivial equations are those for $C^{(3)},C^{(1)}$, and the $v$-$v$ component of the metric.


The presence of internal gauge modes generalizes the fivebrane effective action to \cite{Eyras:1998hn} 
\begin{align}
\label{GeneralAction}
\cS_{\NSfive} &=-T_{A5}\int\! d^6 \xi\, e^{-2\Phi}\sqrt{ |\text{det}\left(G - e^{2\Phi}\mathcal{G} ^{(1)} \mathcal{G} ^{(1)}\right)|}
\left(1+\frac{1}{4!} e^{2\Phi} (\mathcal{H} ^{(3)})^2+\dots\right)-T_{A5} \int_{\cW} \widetilde{\cF} ~.
\end{align}
The ellipses refer to quartic and higher-order terms in $\mathcal{H}$, which as we explain below, are not contributing to the classical equations of half-BPS fivebranes. 
Again, all the bulk fields in the spacetime $\cM$ are pulled back to the fivebrane worldvolume $\cW$ by the fivebrane embedding $\sfF:\cW\to\cM$, \eg\ $\sfF^*(G), \sfF^*(B)$, \etc; we will leave this pullback implicit in what follows.  We will also eventually take the worldvolume coordinates to match with the coordinates $u,v = t\pm \ytil$ as well as the coordinates along $\bT^4$. 
The worldvolume one-form $\cG$ is defined by
\begin{equation}
\label{curlyG}
\mathcal{G} ^{(1)}=C^{(1)}+d \mfa^{(0)}~, 
\end{equation}
where $\mfa^{(0)}$ is the scalar in the IIA fivebrane tensor multiplet; the worldvolume three-form
\begin{equation}
     \label{curlyH}
	\mathcal{H}^{(3)} =  C^{(3)}+ d\mfa^{(2)}+\,d \mfa^{(0)}\!\wedge\! B 
\end{equation}
is the self-dual field strength (with $\mfa^{(2)}$ the chiral tensor potential); 
and the 6-form $\widetilde{\cF}^{(6)}$ is the Wess-Zumino term associated to the 6-form potential $\widetilde B_6$:
\begin{equation}
\label{WZ}
\mathcal{\widetilde{F}} ^{(6)}=\widetilde{B}^{(6)}+d\mfa^{(0)}\wedge C^{(5)}-\frac{1}{2} C^{(3)}\wedge \,d \mfa^{(2)}\! -\frac{1}{2} B^{(2)} \wedge d\mfa^{(0)}\wedge d\mfa^{(2)} ~.
\end{equation}
There are a few technical differences between the action written in reference \cite{Eyras:1998hn} and Eq.~(\ref{GeneralAction}). First, the worldvolume gauge modes in our convention $\mfa_{our}=(2\pi \alpha')^{-1} \times \mfa_{their}$.
Second, a ``dynamical tension'' degree of freedom and a term proportional to $B^{(2)}\wedge \,C^{(3)}\!\wedge d \mfa^{(0)}$ do not exist in our formalism. 
Third, we write the opposite signs of the kinetic term $\mathcal{H}^2$ and some of the Wess-Zumino terms relative to \cite{Eyras:1998hn} (the signs are identical for $\widetilde{B}_6$ and $d\mfa^{(0)} \wedge C^{(5)}$). This is important for finding the supergravity solution and for the gauge invariance of the bulk plus brane action.

Neither the bulk supergravity nor the fivebrane effective actions are individually invariant under gauge transformations involving $C^{(3)}$ and $\mfa^{(2)}$, 
\begin{equation}
\label{Gauge}
	C_3 \to C_3 + d\Lambda_2 ~~,~~~~ \mfa_2 \to \mfa_2 - \Lambda_2 ~,
\end{equation}
rather their sum is gauge invariant: The variation of the Chern-Simons term $\cI_{CS}$ is
\begin{align}
\label{Variation}
		&\delta_{\Lambda_2} \cI_{CS} =		\frac{1}{4\kappa_{0}^2} \int H_3 \wedge d\Lambda_2 \wedge F_4 =
        -\frac{1}{4\kappa_{0}^2} \int dH_3 \wedge \Lambda_2 \wedge F_4-\frac{1}{4\kappa_{0}^2} \int H_3 \wedge \Lambda_2 \wedge dF_4~. 
\end{align}
(The asymptotic boundary term $\frac{1}{4\kappa_{10}^2}\int_{\partial M} H_3 \wedge \Lambda_2 \wedge F_4$ vanishes on the solutions~\eqref{gen STube NS}-\eqref{HpRR} considered here.)
For the fivebrane source, one has
\begin{equation}
\label{dH}
	dH_3 = 2\kappa_0 ^2 T_{A5}\, dx_1 \wedge dx_2 \wedge dx_3 \wedge dx_4 \,\delta^4 (\vec{x} -\vec{\sfF}(v)) \neq 0~,
	\end{equation}
and also
\begin{equation}
\label{dF}
     dF = -dv \wedge dx_1 \wedge dx_2 \wedge dx_3 \wedge dx_4  \nabla^2 \sfZ_0 = -2\kappa_0 ^2 \tau_{\NSfive}\,\partial_v \mfa^{(0)} \,\delta^4 (\vec{x}-\vec{\sfF}(v)) dv \wedge dx_1 \wedge dx_2 \wedge dx_3 \wedge dx_4 ~.
    \end{equation}
The last equality will be derived in Eq.~(\ref{Z0Eq}).
Plugging Eq.~(\ref{dH}) in Eq.~(\ref{Variation}) results in
\begin{equation}
	\delta_{\Lambda_2} \cI_{CS} = -\frac{T_{A5}}{2} \int_{\cW}  \Lambda_2 \wedge F_4+\frac{T_{A5}}{2} \int_{\cW} H_3 \wedge \Lambda_2 \wedge d \mfa^{(0)} ~. 
\end{equation}
On the other hand, the gauge transformation (\ref{Gauge}) vary  $\mathcal{S}_{\NSfive}$ due to two of the terms in the Wess-Zumino six-form $\widetilde{\mathcal{F}}$ as written in 
\begin{equation}
	\delta_{\Lambda_2} \mathcal{S}_{\NSfive}= \frac{T_{A5}}{2}\int_{\cW}  \delta_{\Lambda_2}\left[ \Big(C^{(3)}+ B^{(2)}\wedge d\mfa^{(0)}\Big) \wedge d\mfa^{(2)}\right] = \frac{T_{A5}}{2}  \int_{\cW} \left( \Lambda_{2} \wedge F_4 - H_3 \wedge \Lambda_2 \wedge d\mfa^{(0)}\right)~.
\end{equation}
Thus the combined bulk plus brane effective action is gauge invariant.

We explain the part of the action (\ref{GeneralAction}) associated to the internal scalar $\mfa^{(0)}$.
From an 11d M-theory perspective, $\mfa^{(0)}$ is just a fifth transverse scalar component, which specifies where the fivebrane (now an M5) is in the eleventh dimension:
\begin{equation}
	\sfF^{11}  = \mfa^{(0)}~.
\end{equation}
We will drop the superscript on this scalar when the context is unambiguous (similarly for the two-form potential $\mfa^{(2)}$).
The 11-dimensional line element is written 
\begin{equation}
\label{11d metric}
	ds^2 _{11} = e^{-\frac{2\Phi}{3}} \Big(  ds_{10}^2 + e^{2\Phi} \big(dx_{11} - C_{\mu} dx^{\mu}\big)^2\Big)~.
\end{equation}
The pullback of the target-space metric thus receives an additional contribution, which we write in
\begin{align}
\label{11d pullback}
G_{ab} &= G ^{(10)}_{\mu \nu}\frac{\partial \sfF^{\mu}}{ \partial \sigma^a} \frac{\partial \sfF^{\nu}}{\partial \sigma ^b} 
+e^{2\Phi}\bigg(C _{\mu} \frac{\partial \sfF^{\mu}}{ \partial \sigma^a} -\frac{\partial\mfa}{\partial\sigma^a}\bigg)\bigg(C _{\nu} \frac{\partial \sfF^{\nu}}{ \partial \sigma^b} -\frac{\partial\mfa}{\partial\sigma^b}\bigg)
~.
\end{align}

After compactifying on the $\bT^4$ and introducing the intrinsic metric on the worldsheet as an extra variable, the source effective action again contains a Polyakov-like term and a Wess-Zumino term for the magnetic $\widetilde{B}$-field
\begin{equation}
\label{PolyScalar}
    \cS_{1} = -\frac{\tau^{~}_{\rm\sst NS5}}{2} \int d^2 \sigma \left( e^{-2\Phi} \sqrt{-\gamma} \gamma^{ab} G _{ab} +\epsilon^{ab} \widetilde{B}_{ab} \right)~, 
\end{equation}
where the internal scalar excitations and their coupling to the R-R one-form are incorporated through the pullback of the M-theory metric~\eqref{11d pullback}

The dependence of the M5-brane effective action on the internal antisymmetric gauge field unifies the coupling to $B$ and to $C^{(3)}$ in the field strength $\cH$~\eqref{curlyH}.  The dependence of the action on $\mfa^{(2)}$ is more complicated than for $\mfa^{(0)}$.  Fortunately, for 1/2-BPS solutions we will only need this dependence to quadratic order~-- higher orders lead to equations of motion containing at least one power of $\cH_{uab}$, which vanishes on-shell.  These higher order terms are the ellipses in~\eqref{GeneralAction}, which we will henceforth suppress.  

There are subtleties arising from the fact that the worldvolume tensor gauge field is chiral~\rcite{Witten96}.   
For convenience, let us introduce a basis of self-dual and anti-self dual two-forms on $\bT^4$:%
\footnote{If instead the fivebrane is compactified on $K3$, there will be 16 additional harmonic two-forms, related to the duality between type IIA on $K3$ and the Heterotic string on $\bT^4$~\rcite{Harvey:1995rn}.  Overall, there are 20 internal bosonic R-R modes in addition to the four transverse modes of the fivebrane, and no fermion modes. }
\begin{align}
\label{2forms}
    &\Omega_{(1)} = \frac{1}{\sqrt{2}} (dz_1 \wedge dz_2 + dz_3 \wedge dz_4 ) ~~,~~~~ \widetilde{\Omega}_{(\tilde{1})} = \frac{1}{\sqrt{2}} (dz_1 \wedge dz_2 - dz_3 \wedge dz_4 )
    \nonumber\\
    & \Omega_{(2)} = \frac{1}{\sqrt{2}} (dz_1 \wedge dz_3 + dz_4 \wedge dz_2 ) ~~,~~~~ \widetilde{\Omega}_{(\tilde{2})} = \frac{1}{\sqrt{2}} (dz_1 \wedge dz_3 - dz_4 \wedge dz_2 )
    \\
    &  \Omega_{(3)} = \frac{1}{\sqrt{2}} (dz_1 \wedge dz_4 + dz_2 \wedge dz_3 ) ~~,~~~~ \widetilde{\Omega}_{(\tilde{3})} = \frac{1}{\sqrt{2}} (dz_1 \wedge dz_4 - dz_2 \wedge dz_3 )~.
    \nn
\end{align} 
Then two-forms on $\bT^4$ admit an expansion in terms of these forms, for instance for the tensor gauge field
\begin{equation}
\label{twoform}
	\mfa^{(2)} = \sum_{\asymlabel=1} ^3 a_{\asym} \Omega_{\asymlabel} ^{(2)} + \sum_{\tilde{\asymlabel}=1} ^{3} \tilde{a}_{\asymtil} \widetilde{\Omega}_{\asymlabel} ^{(2)}~.
\end{equation}
In reducing the worldvolume theory to two dimensions, 6d chirality correlates 2d chirality with $\bT^4$ chirality; $a_\asym$ are left-moving ($v$-dependent), while $\tilde a_\asymtil$ are right-moving ($u$-dependent).
The difficulties associated to writing a covariant Lagrangian formulation of chiral two-forms in 6d become difficulties with chiral scalars in 2d.

Three-forms with two legs on the torus also admit such an expansion (let $b$ label a worldvolume dimension transverse to the torus):
\begin{equation}
	C^{(3)} _ b = \sum_{\asymlabel=1} ^3 C_{b\asym} \Omega_{\asymlabel} ^{(2)} + \sum_{\tilde{\asymlabel}=1} ^{3} \widetilde{C}_{b\asymtil} \widetilde{\Omega}_{\asymlabel} ^{(2)}.
\end{equation}

The fivebrane effective action simplifies for half-BPS states. 
In the kinetic term $\cH^2$ in Eq.~(\ref{GeneralAction}), the contribution $\partial\mfa ^{(0)}\wedge B_2$ vanishes for the 1/2-BPS fivebrane solutions we are considering because $\mfa=\mfa(v)$, and so the relevant components in the pullback of $B^{(2)}$ must have legs on the $u$ and/or torus directions. However, such $B^{(2)}$ field components vanish on shell for the ansatz (\ref{gen STube NS}). Moreover, the variations of the $\cH^2$ kinetic term with respect to all $B$-field components vanish as well: 
\begin{equation}
    \frac{\delta }{\delta B _{\mu \nu}(\sfx)}  \int d^6 \xi ~\mathcal{H}^2= 2 G^{vv} G^{aa'} G^{bb'} \mathcal{H}_{vab} \frac{\partial \sfF^{\mu}}{\partial \xi^{a'}}\frac{\partial \sfF^{\nu}}{\partial \xi^{b'}}  \partial_v \mfa^{0}\, \delta^4 (\sfx-\sfF(v))=0~,
\end{equation}
since $G^{vv}=0$ on shell.  The gauge-invariant three-form field strength thus simplifies to
\begin{equation}
     \mathcal{H} = C^{(3)} +  d\mfa^{(2)}~.
\end{equation}
We verify after the fact that this field strength is self-dual for the solutions we consider; the solution (\ref{gen STube NS})-(\ref{RRSources}) admits a self-dual field strength $\mathcal{H}$~-- any $u$ or $\asymtil$ leg makes it zero, and the non-trivial component is
\begin{equation}
    \mathcal{H} _{v \asym} = (*\mathcal{H}) _{v \asym}
\end{equation}

Using the expansion~\eqref{2forms}, \eqref{twoform} the $\cH^2$ kinetic term reduces on the $\bT^4$ as follows
\begin{align}
\label{H KEterm}
\cS_2 = 
\,\,&-\frac{\tau_{\NSfive}}{2} \int d^2 \sigma \sqrt{-\gamma} \gamma^{ab} \sum_{\asymlabel=1} ^3 \left( C_{a(\asymlabel)} + \partial_a \mfa^{(2)} _{(\asymlabel)}\right) \left(C_{b (\asymlabel)} + \partial_b \mfa^{(2)} _{(\asymlabel)}\right)
\nonumber\\
\,\,&-\frac{\tau_{\NSfive}}{2} \int d^2 \sigma \sqrt{-\gamma} \gamma^{ab} \sum_{\tilde{\asymlabel}=1} ^3 \left( \widetilde{C}_{a(\tilde{\asymlabel})} + \partial_a \tilde{\mfa}^{(2)} _{(\tilde{\asymlabel})}\right) \left(\widetilde{C}_{b (\tilde{\asymlabel})} + \partial_b \tilde{\mfa}^{(2)} _{(\tilde{\asymlabel})}\right)~.
\end{align}
We will consider the variation of the $\cH^2$ kinetic term with respect to $C^{(3)}$ below.  Let us instead complete the torus reduction of the fivebrane effective action by considering the various contributions to the Wess-Zumino term (\ref{WZ}) in~(\ref{GeneralAction}).  The term
\begin{equation}
      \mathcal{\widetilde{F}}^{(6)}\supset B\wedge  d \mfa^{(2)} \wedge  d \mfa^{(0)}  
\end{equation}
has a vanishing variation of with respect to $B_{\mu \nu}$ because all derivatives must be $v$-derivatives. 
The term 
\begin{equation}
    \mathcal{\widetilde{F}}^{(6)}\supset C^{(5)}\wedge d \mfa^{(0)}
\end{equation}
sources $D4$ dipole charge; in a democratic formalism for supergravity one would write an equation of motion for $C^{(5)}$ that has a source proportional to $\sum_{\I=1} ^{n_5} *_6\partial\mfa_\I^{(0)} \delta^4 \big(\sfx - \sfF(v)\big)$. However, we have chosen to work in the standard formalism of supergravity where a Hodge duality relation ties $C^{(5)}$ with $C^{(3)}$: 
\begin{equation}
\label{HodgeDuality}
    \widetilde{F}^{(6)}=* \widetilde{F}^{(4)}~,
\end{equation}
where $\widetilde{F}^{(q)} = dC^{(q-1)} -H^{(3)} \wedge C^{(q-3)}$. 
This means that the $D4$ flux can be obtained from the $D2$ flux.
To summarize, in the standard formalism of supergravity, the Wess-Zumino term for 1/2-BPS NS5-branes boils down to the $\widetilde B$ term in~\eqref{PolyScalar}, together with a term on the effective string
\be
\label{a2 WZterm}
\cS_3 = -\frac{\tau_{\NSfive}}{2}\int d^2 \sigma \, \epsilon^{ab} \Big( \sum_{\asymlabel=1} ^3  \partial_a a^{(2)} _{(\asymlabel)} C^{(3)} _{b(\asymlabel)}- \sum_{\tilde{\asymlabel}=1} ^3\partial_a \tilde{a}^{(2)} _{(\tilde{\asymlabel})} \widetilde{C}^{(3)} _{b(\tilde{\asymlabel})} \Big)~.
\ee
We again choose the conformal gauge on the effective string, and as in the discussion in the previous subsection, the constraints and reparametrization invariance lead us to set $\partial_u\sfF^u=1$.  

The sum of $\cS_3$ in Eq.~(\ref{a2 WZterm}) and $\cS_2$ in Eq.~(\ref{H KEterm}) is given by the simple formula
\begin{align}
\label{SimplifiedS23}
    \cS_2+\cS_3=
    \frac{\tau_{\NSfive}}{2} \int d^2 \sigma  \bigg[ &\sum_{\asymlabel=1} ^3\, \Big(  C_{u (\asymlabel)} C_{v(\asymlabel)} +2 C_{u(\asymlabel)} \partial_v \mfa^{(2)} _{(\asymlabel)}+\partial_u \mfa^{(2)} _{(\asymlabel)} \partial_v \mfa^{(2)} _{(\asymlabel)}\Big) \nn \\
    +&  \sum_{\tilde{\asymlabel}=1} ^3 \Big( \widetilde{C}_{u (\tilde{\asymlabel})} \widetilde{C}_{v(\tilde{\asymlabel})} +  2\widetilde{C}_{v (\tilde{\asymlabel})} \partial_u \tilde{\mfa}^{(2)} _{(\tilde{\asymlabel})}+\partial_u \tilde{\mfa}^{(2)} _{(\tilde{\asymlabel})} \partial_v \tilde{\mfa}^{(2)} _{(\tilde{\asymlabel})}\Big)\bigg]~.
\end{align}
A partial summary is in order. The kinetic term in Eq.~(\ref{H KEterm}) contains four pieces of the two-form potential: left/right-moving along $\bS^1_\ytil$, and self/antiself-dual on $\bT^4$. The addition of the Wess-Zumino term in Eq.~(\ref{a2 WZterm}) has caused the cancellation of two unwanted terms proportional to $C_{v \asym} \partial_u \mfa^{\asym}$ and $\widetilde{C}_{u \asym} \partial_v \tilde{\mfa}^{\asymtil} $, so that the left-moving, anti-self-dual part and the right-moving, self-dual part of $\mfa^{(2)}$ do not couple to the bulk field $C^{(3)}$. 
The last line of Eq.~(\ref{SimplifiedS23}) involves the right-moving component of the chiral two-form; since we are considering 1/2-BPS solutions which are invariant under $u$ translations, for all practical purposes we are left with the first line of Eq.~(\ref{SimplifiedS23}).%
\footnote{These fields do have equations of motion, because there is no covariant effective action for the chiral two-form by itself.  We follow the usual strategy of using the action involving them to derive the equations of motion for $\mfa^{(2)}$, and then at that point imposing the self-duality condition on $\cH$.}
Altogether, the source effective action reduces to
\be
\cS_{\rm source} = \cS_1+\cS_2+\cS_3  ~.
\ee
We are now prepared to derive the general solution~\eqref{gen STube NS}-\eqref{HpRR}, \eqref{H5result}, \eqref{Aresult}.

\paragraph{Equations for $C^{(3)}$:}
The calculation of the equation following from varying relative to $C^{(3)} _{u (\tilde{\asymlabel})}$ brings a trivial equation, however, the variation relative to $C^{(3)} _{u (\asymlabel)}$ produces a non-trivial result.
Varying the source effective action in Eq.~(\ref{GeneralAction}) gives 
\begin{align}
\label{deltaIdeltaC3}
	 &\frac{\delta S_{\rm source}}{\delta C_{u \asym} ^{(3)}}= \frac{\tau_{\NSfive}}{12} \big( C_{ v \asym}  
 +2 \partial_{v} \mfa_{\asym} \big) ~.
\end{align}
Varying the entire action with respect to $C^{(3)} _{u\asym}$ leads to
\begin{align}
\label{C3nontrivialEq} 
\partial_{i} \left(\sqrt{-G} G^{ii}G^{uv}\widetilde{F}_{i v (\asymlabel) }\right) & +\frac{1}{2\times 4!}\epsilon^{ijkl} (dH)_{ijkl} C_{v(\asymlabel)}+\frac{1}{4!}\epsilon^{ijkl} H_{ijk}F^{(4)} _{lv (\asymlabel)}
\nn\\
&\hskip 2.5cm
=-\kappa_0^2 \tau_{\NSfive} \left( C_{v (\asymlabel)} + 2\partial_v \mfa_{(\asymlabel)}\right)  \delta^4 (\sfx-\sfF(v))~.
\end{align}
where $\epsilon^{ijkl}$ is the Levi-Civita tensor on the transverse $\bR^4$.
We plug
\begin{align}
	F_{iv \asym}=\widetilde{F}_{iv \asym} &= \partial_i \bigg(\frac{\sfZ_\asym}{\sfH_5}\bigg)~, (dH)_{ijkl} = -\nabla^2 H_5 \epsilon_{ijkl}  
\end{align}
into Eq.~(\ref{C3nontrivialEq}). The first term on the L.H.S is given by
\begin{align}
-\partial_i \bigg( \sfH_5\, \partial_i \Big(\frac{\sfZ_{\asym } }{\sfH_5}\Big) \bigg)
&=-\partial_i \left( \partial_i \sfZ_{\asym } -\frac{1}{\sfH_5} \sfZ_{\asym }  \partial_i \sfH_5  \right)
 \nonumber\\
	& =
  -\nabla^2 \sfZ_{\asym }  +\frac{1}{\sfH_5} \sfZ_{\asym } \nabla^2 \sfH_5 -  \frac{\partial_i \sfH_5 \partial_i \sfH_5}{\sfH_5 ^{\,2}}\sfZ_{\asym } +\frac{1}{\sfH_5} \partial_i \sfZ_{\asym }  \partial_i \sfH_5 ~.
\end{align}
The second term on the L.H.S of Eq.~(\ref{C3nontrivialEq}) is
\begin{align}
 \frac{1}{2\times 4!} \epsilon^{ijkl} (dH)_{ijkl} C_{v \asym } = \kappa_0 ^2  \tau_{\NSfive}  \delta^4 (\vec{x}-\vec{\sfF}(v)) \frac{Z_{\asym}}{\sfH_5}~.
\end{align}
The third term on the L.H.S of equation (\ref{C3nontrivialEq}) is 
\begin{align}
& \frac{1}{4!}\epsilon^{ijkl }H_{ijk} F^{(4)} _{lv (\asymlabel)} =  -\partial_l \sfH_5 \, \partial_l \left( \frac{\sfZ_\asym}{\sfH_5}  \right)
= \frac{ \partial_l \sfH_5\, \partial_l \sfH_5 }{\sfH_5 ^{\,2}}\,\sfZ _{\asym}-\frac{1}{\sfH_5}\partial_l \sfZ_{\asym } \partial_l \sfH_5\,   ~.
\end{align}
It follows that
\begin{equation}
	\nabla^2 \sfZ_\asym = 2\kappa_0^2 \tau_{\NSfive} \sum_{\I=1 } ^{n_5}\partial_v \mfa _{\I}^{(\asymlabel)} \delta^4 (\sfx - \sfF_\I (v))~,
\end{equation}
which implies
\begin{equation}
\label{ZT4Result}
	\sfZ_\asym (v,\sfx) = -\sum_{\I=1}  ^{n_5}\frac{\alpha' \partial_v \mfa _\I^{(\asymlabel)} (v)}{|\sfx-\sfF _\I (v)|^2}~.
\end{equation}

In the analysis of the $C^{(1)}$ equations of motion, we will need the field strength $\widetilde{F}_{vijk}$, which we obtain through a variation of the action with respect to $C^{(3)}_{uij}$; this leads to the equation of motion 
\begin{equation}
 \partial_{\mu} \left( \sqrt{-G} \,\widetilde{F}^{\mu uij} \right)=\epsilon ^{ijkl }H_{vkl}F _{(\asymlabel) (\asymlabel)}=0~. 
\end{equation}
Therefore,
\begin{equation}
    \widetilde{F}_{vijk} = \sfH_5 \,\epsilon_{ijkl} \partial_l \mathcal{E}~,
\end{equation}
for an arbitrary scalar function $\mathcal{E}$. The value of this function that will be consistent with the solution is
\begin{equation}
\label{Fvijk}
\mathcal{E} = \frac{\sfZ_0}{\sfH_5} 
~~\Longrightarrow~~ 
F_{vijk} = \epsilon_{ijkl} \partial_l \sfZ_0~.
\end{equation}
The variation of the action with respect to all the other components of $C^{(3)}$ lead to trivial equations of motion.

Note also that the Hodge duality relation in Eq.~(\ref{HodgeDuality}) is satisfied for
\begin{align}
C^{(5)} &= \frac{\sfZ_0}{\sfH_5} dv \wedge \widehat{\text{vol}}_4+  \mathbf{c}_{\asym}\wedge dv\wedge \Omega_\asym ~,
\end{align}
where
\begin{equation}
    d\mathbf{c} = *_4 d\sfZ_0 ~~,~~~~ d\mathbf{c}_{\asym} = *_4 d\sfZ_{\asym}   ~. 
\end{equation}
\paragraph{Equations for $C^{(1)}$:}
Trivial equations occur when taking the variation of the action with respect to $C_{\mu} ^{(1)}$ with $\mu\neq u$. 
Varying the term in the effective action $\cS_1$ Eq.~(\ref{PolyScalar}) with respect to the 1-form R-R potential affects the contribution to the pullback metric Eq.~(\ref{11d pullback}) and this gives
\begin{equation}
\label{Polchy}
\frac{\delta S_{\rm source}}{\delta C_{\mu}} = -\tau_{\NSfive}  \sqrt{-\gamma} \gamma^{ab} \left( C_{\nu} \frac{\partial \sfF^{\nu}}{\partial \sigma^b} - \partial_b \mfa^{(0)} \right) \frac{\partial \sfF^{\mu}}{\partial \sigma^a}   \delta^4 (\sfx-\sfF(v))~.
\end{equation}
For $\mu=u$ we have
\begin{equation}
\label{C1v}
\frac{\delta S_{\rm source}}{\delta C_{u}} = \tau_{\NSfive}   \left( C_{v}  - \partial_v \mfa^{(0)} \right)   \delta^4 (\sfx-\sfF(v))~.
\end{equation}
The equation of motion~\eqref{RR1} for the R-R 1-form potential is thus
\begin{align}
	-\frac{1}{2\kappa_0^2 \sqrt{-G}} \partial_{\mu} \Big(\sqrt{-G} & F^{\mu \nu}\Big)+\frac{1}{12\kappa_0^2} H_{\alpha \beta \gamma} \left( F ^{\nu \alpha \beta \gamma } + C^{\nu} H^{\alpha \beta \gamma} - 3 C^{\alpha} H^{\nu \beta \gamma} \right)\nonumber\\
& =  \frac{1}{\sqrt{-G}}\tau_{\NSfive} \sum_{\I=1} ^{n_5} \left( C_{v}  - \partial_v \mfa^{(0)} _\I\right)  \delta^4 \left(\sfx-\sfF_\I(v)\right)~.
\end{align}
Then
\begin{align}
	\label{C^u}
	-\frac{1}{2\kappa_0^2 \sqrt{-G}} \partial_v & \left( \sqrt{-G} F^{vu}\right) 	-\frac{1}{2\kappa_0 ^2 \sqrt{-G}} \partial_i\left( \sqrt{-G}F^{iu}\right)\nonumber\\
	\hskip 2cm
 &=-\frac{1}{12 \kappa_0^2} \left( F^{u ijk} H_{ijk} +3F^{uvij} H_{vij}+C^{u} H_{ijk} H^{ijk}-9C^{v}H_{vij} H^{uij}-3C^{i} H_{ijk} H^{ujk} \right)\nonumber\\
	&\hskip 2cm +\frac{1}{\sqrt{-G}}\tau_{\NSfive} \sum_{\I=1} ^{n_5}   \left(C_v  -  \partial_v \mfa^{(0)} _\I\right) \delta^4 (\sfx-\sfF_\I(v))~.
\end{align} 
Many of the terms on the R.H.S of Eq.~(\ref{C^u}) vanish. A non-trivial contribution to the L.H.S arises from
\begin{align}
	F^{iu} &= G^{ii} G^{uv}  F_{iv}=-\frac{2}{\sfH_5} \partial_i C_v~.
\end{align}
Substituting these equations in (\ref{C^u}) implies
\begin{align}
	\label{C^u2}
	\frac{1}{\kappa_0 ^2 \sfH_5 ^{\,2}} \partial_i \left(\sfH_5  \partial_i C_v\right) &=\frac{1}{6 \kappa_0 ^2} C_v H_{ijk} H^{ijk}-\frac{1}{12 \kappa_0^2}  G^{uv} G^{ii} G^{jj} G^{kk} F_{vijk} H_{vij}\nonumber\\
	&\hskip 2cm +\frac{2}{H_5 ^2}\tau_{\NSfive} \sum_{\I=1} ^{n_5}   \left( C_v  - \partial_v \mfa^{(0)} _\I\right) \delta^4 (\sfx-\sfF_\I(v))~.
\end{align}
For
\begin{equation}
C_v(v,\sfx) = -\frac{\sfZ_0 (v,\sfx)}{ \sfH_5 (v,\sfx)}~,
\end{equation}
we can compute
\begin{align}
	\frac{1}{\kappa_0^2 \sfH_5 ^{\,2}} \partial_i \left(\sfH_5 \partial_i C_v\right) &=-\frac{1}{\kappa_0^2  \sfH_5 ^{\,2}} \partial_i \left(\sfH_5  \partial_i \frac{\sfZ_0}{\sfH_5}\right) =\frac{1}{\kappa_0 ^2 \sfH_5 ^{\,2}} \partial_i \left(-\partial_i \sfZ_0 + \frac{\sfZ_0}{\sfH_5} \partial_i \sfH_5\right)
\nonumber\\[.2cm]
&= -\frac{1}{\kappa_0 ^2 \sfH_5 ^{\,2}} \left( \nabla^2 \sfZ_0 + C_v \nabla^2 \sfH_5 - \frac{C_v}{\sfH_5} \partial_i \sfH_5 \partial_i \sfH_5 - \frac{\partial_i \sfZ_0 \partial_i \sfH_5}{\sfH_5} \right)~.
\end{align}
The R.H.S of Eq.~(\ref{C^u2}) contains
 \begin{equation}
	\frac{1}{6\kappa_0^2} C_v H_{ijk} H^{ijk} = \frac{C_v}{6\kappa_0^2 \sfH_5 ^{\,3}} (\epsilon_{ijkl} \partial_l \sfH_5)^2=\frac{C_v}{\kappa_0^2 \sfH_5 ^{\,3}} (\partial_i \sfH_5)^2~.
\end{equation}
as well as
\begin{equation}
	\sum_{\I=1} ^{n_5}\frac{2\tau_{\NSfive}}{\sfH_5 ^2}  C_v \,\delta^{(4)} (\sfx-\sfF_\I(v)) = - \frac{1}{\kappa_0^2 \sfH_5 ^{\,2}} C_v \nabla^2 \sfH_5~.
\end{equation}
We now use Eq.~(\ref{Fvijk}) which allow one to proceed to
\begin{align}
	-\frac{1}{12 \kappa_0^2} G^{uv} G^{ii} G^{jj} G^{kk} F_{vijk}H_{ijk}=\frac{1}{6\kappa_0^2 \sfH_5 ^{\,3} } \epsilon_{ijkl} \partial_l \sfZ_0 \epsilon_{ijkl} \partial_l \sfH_5 =\frac{1}{\kappa_0^2  \sfH_5^ {\,3}} \partial_i \sfZ_0 \partial_i \sfH_5 ~.
\end{align}
Combining the equations above, we obtain
\begin{equation}
\label{Z0Eq}
\nabla^2 \sfZ_0 = 2\kappa_0 ^2 \tau_{NS5} \sum_{\I=1} ^{n_5}\partial_v  \mfa_\I ^{(0)} \delta^4 \left(\sfx -\sfF_{\I}(v)\right) ~,
\end{equation}
and thus
\begin{equation}
\label{Z0Result}
\sfZ_0 (v,\sfx) = -\alpha' \sum_{\I=1} ^{n_5}\frac{ \partial_v \mfa_\I ^{(0)}}{\left|\sfx - \sfF_\I(v)\right|^2}~.
\end{equation}

\paragraph{Modification of $\sfH_p$:}  
Next, we want to write the $v$-$v$ component of the Einstein equation. As in the previous subsection, it is useful to write the constraint equation coming from the variation of the action with respect to the inverse intrinsic metric on the worldsheet. When gauge modes on the fivebranes are excited, from Eqs.~(\ref{PolyScalar}) and (\ref{H KEterm}) we find 
\begin{align}
\label{Newconstraint} 
&\sum_{\I=1} ^{n_5} \left( e^{-2\Phi}G_{ab} + \left( C_a ^{(1)} -\partial_a \mfa^{(0)} _{\I}\right)\left( C_b ^{(1)} -\partial_b \mfa^{(0)} _{\I}\right)+  \left( C^{(3)}_{a\asym} +\partial_a \mfa^{(2)} _{\I \asym}\right)\left( C_{b\asym } ^{(3)} +\partial_b \mfa^{(2)} _{\I \asym}\right) \right)=\nonumber\\
&\hskip .5cm
    \frac{1}{2}~\gamma_{ab}\gamma^{cd}\sum_{\I=1} ^{n_5} \left( e^{-2\Phi}G_{cd} + \left( C_c ^{(1)} -\partial_c \mfa^{(0)} _{\I}\right)\left( C_d ^{(1)} -\partial_d \mfa^{(0)} _{\I}\right)+  \left( C ^{(3)}_{c\asym} +\partial_c \mfa^{(2)} _{\I \asym}\right)\left( C ^{(3)}_{d\asym} +\partial_d \mfa^{(2)} _{\I \asym}\right) \right)~.
\end{align}
For $a,b=u$ Eq.~(\ref{Newconstraint}) is trivially satisfied in the conformal gauge $\gamma_{uu}=0$. For $a=v,b=u$ and $a=u,b=v$, Eq.~(\ref{Newconstraint}) is satisfied as in the case without internal excitations, because all RR fields and internal excitations with a $u$ leg (or derivative in the case of $\mfa^{(0)}$) in the equation vanish. For $a=b=v$, the R.H.S of Eq.~(\ref{Newconstraint}) vanishes since $\gamma_{vv}=0$ and 
\begin{align}
&\sum_{\I=1} ^{n_5} \bigg( 2G_{uv} \partial_u \sfF^u _{\I} +\sfH_5 \big|\partial_v \vec{\sfF}_{\I} (v)\big|^2 + 2\vec{\sfA} \cdot \partial_v \vec{\sfF}_{\I} (v) + \sfH_P  -\frac{\sfZ_{0} ^2 + \sfZ_{(\ell)}^2 }{\sfH_5}\bigg)+\nonumber\\
&\hskip 3cm 
\sfH_5	\sum_{\I=1} ^{n_5}\bigg[ \bigg( \frac{\sfZ^{(0)} }{\sfH_5}+\partial_v \mfa^{(0)}_{\I} \bigg)^2+\bigg( \frac{\sfZ^{(\ell)} }{\sfH_5}+\partial_v \mfa^{(2)}_{\I(\ell)}\bigg)^2 \bigg]=0~.
\end{align}
Therefore,
\begin{align}
\sum_{\I=1} ^{n_5} 2 G_{uv} \partial_u \sfF^u _{\I} 
&= -\sum_{\I=1} ^{n_5}  \sfH_5 \Big(|\partial_v \vec{\sfF}_{\I} (v)|^2+|\partial_v \mfa^{(0)} _{\I} (v)|^2+|\partial_v \mfa^{(2)} _{\I(\ell)} (v)|^2\Big)\nonumber\\
&\hskip 1cm
- \sum_{\I=1} ^{n_5} \left(2\vec{\sfA} \cdot \partial_v \vec{\sfF}_{\I} (v) + \sfH_P +2 \sfZ ^{(0)} \partial_v \mfa^{(0)} _{\I} (v) +2\sfZ ^{(\ell)} \partial_v \mfa^{(2)} _{\I,(\ell)}(v)\right)~.
\end{align}
It follows that the $v$-$v$ component of the target space stress energy tensor is
\begin{align}
T_{vv} &= G_{uv} \left( 2G_{vv} T^{uv} + 2 G_{vi} T^{ui} + G_{uv} T^{uu} \right)  
\nonumber\\
&= \frac{\tau_{\NSfive}}{H_5 ^2} e^{-2\Phi} \sum_{\I=1} ^{n_5} \bigg[\sfH_5 \Big(\big|\partial_v \vec{\sfF}_{\I} (v)\big|^2+\big|\partial_v \mfa^{(0)} _{\I} (v)\big|^2+\big|\partial_v \mfa^{(2)} _{\I(\ell)} (v)\big|^2\Big)\nonumber\\
&\hskip 1cm 
    -\sfH_P+\frac{(\sfZ^{(\ell)})^2+(\sfZ^{(0)})^2}{\sfH_5}+2 \sfZ ^{(0)} \partial_v \mfa^{(0)} _{\I} (v) +2\sfZ ^{(\ell)} \partial_v \mfa^{(2)} _{\I(\ell)}(v) \bigg]\delta^4 (\vec{x}-\vec{\sfF}_{\I} (v))~.
\end{align}
The Einstein $v$-$v$ equation is given by
\begin{align}
e^{-2\Phi} \cG_{vv}
 -\frac{1}{2} F_{v \alpha} F_{v} ^{~\alpha} -\frac{1}{12} \widetilde{F}_{v\alpha \beta \gamma} \widetilde{F}_{v} ^{~\alpha \beta \gamma}+\frac{1}{8} G_{vv} F_{\alpha \beta} F^{\alpha \beta} +\frac{1}{96} G_{vv} \widetilde{F}_{\alpha \beta \gamma \delta} \widetilde{F}^{\alpha \beta \gamma \delta}=\kappa_0 ^2 e^{-2\Phi} T_{vv}~, 
\end{align}
where $\cG_{vv}$ is the $v-v$ component of the tensor defined in Eq.~\eqref{comb2}.
Because the only nonzero components of these field strengths are $F_{iv},F_{vijk},F_{vi\lambda_1\lambda_2}$, and $G^{vv}=0$, we have $F_{\alpha \beta} F^{\alpha \beta}=0$ as well as $\widetilde{F}_{\alpha \beta \gamma \delta} \widetilde{F}^{\alpha \beta \gamma \delta}=0$. Furthermore, we have explicit expressions for the nonzero field strength components, permitting us to compute
\begin{align}
	-\frac{1}{2} F_{v \alpha} F_{v} ^{~~\alpha} -\frac{1}{12} \widetilde{F}_{v\alpha \beta \gamma} \widetilde{F}_{v} ^{~~\alpha \beta \gamma}= -\frac{1}{\sfH_5}  \partial_i \left(  \frac{\sfZ_0}{\sfH_5}\right)\partial_i \left(  \frac{\sfZ_0}{\sfH_5}\right)-\frac{1}{\sfH_5}  \partial_i \left(  \frac{\sfZ_{\asym}}{\sfH_5}\right)\partial_i \left(  \frac{\sfZ_{\asym}}{\sfH_5}\right)~.
\end{align}
The result for the $v$-$v$ metric equation is then
\begin{align}
& 
 -\frac{1}{2\sfH_5 ^{\,2}} \nabla^2 \sfH_p+\frac{\sfH_p}{2\sfH_5 ^{\,3}} \nabla^2 \sfH_5   -\frac{(\sfZ_0) ^2+  |\sfZ_\asym| ^2}{2\sfH_5 ^4} \nabla^2 \sfH_5 +\frac{\sfZ_0\nabla^2 \sfZ_0+  \sfZ_\asym \nabla^2 \sfZ_\asym }{\sfH_5 ^{\,3}}     
\nonumber\\
&\hskip 1.5cm 
=~ \kappa_0 ^2 \frac{\tau_{\NSfive}}{\sfH_5 ^3}\sum_{m=1} ^{n_5} \bigg[\sfH_5 \Big(\big|\partial_v \vec{\sfF}_m (v)\big|^2+\big|\partial_v \mfa^{(0)} _m (v)\big|^2+\big|\partial_v \mfa^{(2)} _{m(\ell)} (v)\big|^2\Big)\nonumber\\
&\hskip 2.5cm 
-\sfH_P+\frac{(\sfZ_{0})^2+(\sfZ_{(\ell)})^2}{\sfH_5}+2 \sfZ _{0} \partial_v \mfa^{(0)} _m (v) +2\sfZ_{(\ell)} \partial_v \sfa^{(2)} _{m(\ell)}(v) \bigg]\delta^4 (\vec{x}-\vec{\sfF}_m (v))
\end{align}
Several cancellations take place, for example
\begin{equation}
	\left( \frac{\sfH_p}{2\sfH_5 ^{\,3}}-\frac{(\sfZ_{0})^2 + |\sfZ_{\asym}|^2}{2\sfH_5 ^4} \right)\nabla^2 \sfH_5 = -\left( \frac{\sfH_p}{\sfH_5 ^{\,3}}-\frac{(\sfZ_{0})^2 + |\sfZ_{\asym}|^2}{2\sfH_5 ^4}\right)  \kappa_0 ^2 \tau_{\NSfive}\sum_{\I=1} ^{n_5} \delta^4 (\sfx-\sfF_{\I}(v))~.
\end{equation}
Also, Eqs.~(\ref{Z0Result}) and (\ref{ZT4Result}) for $\sfZ_0$ and $\sfZ_\asym$ imply that
\begin{align}
	&  \nabla^2 \sfH_p  =-2\kappa_0 ^2 \tau_{\NSfive} \sum_{\I=1} ^{n_5}\Big(  \big|\partial_v \sfF_\I\big|^2 + \big|\partial_v \mfa_{0}^\I\big|^2 + \big|\partial_v \mfa_{\asym}^\I\big|^2 \Big)  \delta^4 \left(\sfx-\sfF_{\I}(v)\right)~.
\end{align}
Consequently,
\begin{equation}
\label{HpRRincluded}
\sfH_p (v,\sfx) = \alpha'\sum_{\I=1} ^{n_5}  \frac{  \big|\partial_v \sfF_\I\big|^2 + \big|\partial_v \mfa_{0}^\I\big|^2 + \big|\partial_v \mfa_{\asym}^\I\big|^2 }{|\sfx-\sfF_\I(v)|^2}~.
\end{equation}
To summarize, we have verified that the configuration (\ref{gen STube NS}),(\ref{gen STube RR}) solves the equations of motion in the presence of explicit source terms for fivebrane excitations.  
All the fivebrane excitations contribute to the momentum harmonic function on the same footing, and the internal fivebrane modes source the R-R potentials.

For the $\bS^1\times\bT^4$ compactification considered above, the spectrum includes eight left-moving fermions which are superpartners of the eight bosonic modes.  Altogether these 1+1d left-moving profiles have the same accounting as the 1/2-BPS states of a type II fundamental string wrapping the $\bS^1$ of $\bS^1\times\bT^4$, since the NS5-P system lies on the same duality orbit.

Our analysis above suppresses the effects of the fermions. 
 In principle one could include them via the supersymmetrization of the fivebrane source effective action above (see for example~\rcite{Aganagic:1997zq}). Fermions induce dipole terms in the harmonic forms and functions, see~\rcite{Taylor:2005db} for a discussion in the context of the F1-P duality frame.  It would be interesting to calculate the harmonic functions in the presence of fermion condensates.

For a K3 compactification, there are no fermions in the left-moving spectrum of the effective string, but rather 24 bosons in total.  In addition to the four transverse scalars on the 6d effective string, there are 19 internal scalars $\mfa^{(2)}$ arising from the self-dual 2-form cohomology of K3, according to the ansatz~\eqref{twoform}, plus the scalar $\mfa^{(0)}$.  Altogether the spectrum is that of 1/2-BPS modes of the heterotic string wrapping the $\bS^1$ of $\bS^1\times \bT^4$, consistent with the fact that it is on the same U-duality orbit as the type IIA NS5-P system on $\bS^1\times K3$~\rcite{Witten:1995ex,Harvey:1995rn}.

\section{The NS5-P ``ensemble geometry''}
\label{sec:NS5-P ensemble}

The space of solutions~\eqref{NS5-P geom} is parametrized by the profile functions $\sfF(v),\mfa_0(v),\mfa_{\asym}(v)$, and classical solutions form a phase space \rcite{Crnkovic:1986ex}.  The symplectic form on this phase space induced from the supergravity action is the natural one would choose for free scalar fields in 1+1d (this was demonstrated for the transverse excitations in~\rcite{Rychkov:2005ji}, and we will assume that the same holds true for the internal excitations).  Quantizing the phase space of 1/2-BPS solutions is thus tantamount to quantizing the effective string whose coordinates include $\sfF(v),\mfa_0(v),\mfa_{\asym}(v)$ as the bosonic left-movers.  Having this Hilbert space in hand, we are free to consider superpositions of states, or mixed states.
The geometry is linear in the harmonic forms and functions (the R-R fields involve simple ratios of harmonic functions), and so it is natural to regard the geometry sourced by a superposition of states to be given by the superposition of harmonic functions (similarly for mixed states and ensemble averages).

This idea leads to the notion of the geometry sourced by a typical 1/2-BPS NS5-P state.  Picking some random element of the ensemble yields a hopelessly complicated source configuration and ambient geometry; however performing the ensemble average leads to a rather simple ``ensemble geometry'' \rcite{Alday:2006nd,Balasubramanian:2008da,Raju:2018xue,Martinec:2023xvf}.
The maximally mixed ensemble equipartitions the excitation budget among all the left-moving scalars (and fermions) on the effective string.  We will find it convenient, however, to treat the ensembles of transverse and internal excitations separately, with independent chemical potentials $2\pi \tauperp$, $2\pi \taupar$, respectively.  This allows us to control separately the excitation levels $N_\perp,N_\parallel$ in the microcanonical ensemble, subject to the constraint 
\be
N=n_5n_p=N_\perp+N_\parallel ~.
\ee

The computation of the ensemble average follows along the same lines as~\rcite{Martinec:2023xvf}, where only the excitations of the transverse scalars were considered (\ie~$N_\parallel$ was set to zero).  The Hilbert space trace is equivalent to a torus path integral over the scalars on the effective string, with the modular parameter $\tau$ the chemical potential conjugate to the momentum of left-moving excitations on $\bS^1_y$.

When the internal profiles $\mfa^{(0)},\mfa^{(2)}$ are included, the R-R harmonic functions $\sfZ_0,\sfZ_\asym$ are linear in the derivatives of these scalars.  The torus one-point function that computes their ensemble average vanishes, in much the same way as the computation for $\llangle\sfA\rrangle$ performed in~\rcite{Martinec:2023xvf}.  The computation of the ensemble average of $\sfH_5$ is the same, with the understanding that the transverse scalars are now at chemical potential $2\pi \tauperp$.  For the harmonic function $\sfH_p$, the calculation of~\rcite{Martinec:2023xvf} factorized into a computation of the expectation value of the stress tensor for $\sfF$ times the same computation that led to $\sfH_5$; for the internal modes, one will trivially obtain their stress tensor as in~\eqref{HpRRincluded} times the harmonic function for $\sfH_5$.  All told, after passing from the canonical to the microcanonical ensemble we have
\begin{align}
\label{NS5P ensemble}
	\llangle \sfH_p (\sfx,v) \rrangle &=\frac{Q_p}{|{\sfx}|^2}\left(1-e^{-\frac{|{\sfx}|^2}{r_b^2}} \right)
\nn\\[.3cm]
	\llangle \sfH_5 (\sfx,v)\rrangle &=\frac{Q_5}{|{\sfx}|^2}\left(1-e^{-\frac{|{\sfx}|^2}{r_b^2}} \right)
\\[.3cm]
	\llangle \sfA(\sfx,v)\rrangle &= \llangle \sfZ_0(\sfx,v)\rrangle  = \llangle \sfZ_\asym(\sfx,v)\rrangle  = 0 ~.
\nn
\end{align}
where we denote the ensemble average by double brackets to distinguish it from an expectation value in a particular Hilbert space state.
The source constitutes a spherical ``blob'' of radius $r_b$, where
\be
r_b = \mu\Big(\frac{\pi^2N_\perp}{6}\Big)^{1/4}  ~,
\ee
where $N_\perp$ is the total excitation level of the transverse scalars $\sfF$.
The overall length scale $\mu$ defined in~\eqref{mudef} is determined by the rescaled inverse tension of the 6d effective string (the fivebrane wrapped on $\bT^4$). 
\begin{equation}
    \tau_{\NSfive} = \frac{1}{2\pi \mu^2} ~~\Longrightarrow~~ \mu^2 = \frac{ (\alpha')^3}{V_{\bT^4}} ~,
\end{equation}
where we have set $\text{Vol}(T^4) = (2\pi)^4 V_{\bT^4}$ and used~\eqref{kappadef}.
The size of the source is easy to understand.  One has a gas of left-movers on the 6d effective string; among these are the modes of the transverse profile which undergoes a random walk~\rcite{Martinec:2023xvf}, with individual steps spanning a thermal wavelength.  There are of order $\sqrt{N_\perp}$ such steps (where $N_\perp$ is the total excitation level of the transverse modes~$\sfF$), leading to a transverse size of order $N_\perp^{1/4}$ in units of the effective string length~$\mu$.

\begin{figure}[ht]
\centering
  \begin{subfigure}[b]{0.5\textwidth}
  \hskip 0cm
\includegraphics[width=\textwidth]{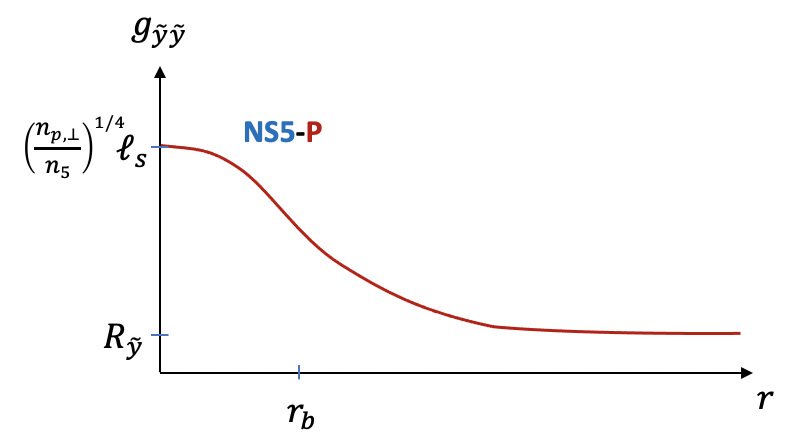}
    \caption{ }
    \label{fig:RYtil}
  \end{subfigure}
\hskip 2cm
  \begin{subfigure}[b]{0.35\textwidth}
  \vskip -2cm
    \includegraphics[width=\textwidth]{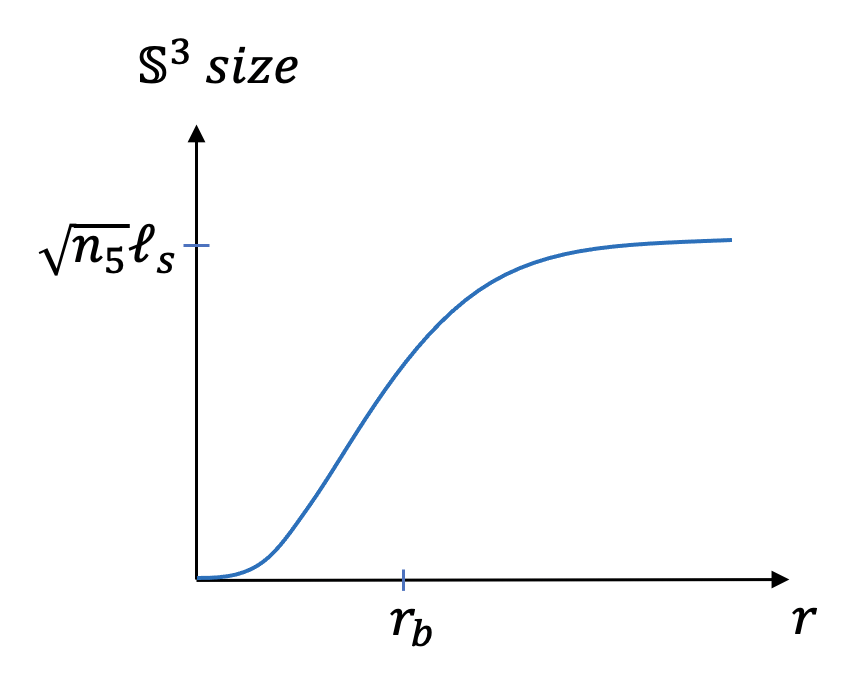}
    \caption{ }
    \label{fig:RS3}
  \end{subfigure}
\caption{ 
\it The NS5-P ensemble geometry in the fivebrane decoupling limit.  The $\tilde y$-circle asymptotes to a fixed size far from the fivebrane source, while close to the source the circle backreacts in response to the momentum and expands to a parametrically larger value.  Meanwhile, as one dives inside the fivebrane blob, the amount of magnetic $H_3$ flux enclosed by a Gaussian sphere decreases, and therefore the proper size of the angular $\bS^3$ shrinks smoothly to zero in the core.
}
\label{fig:RYtil-RS3}
\end{figure}

The characteristic features of the ensemble geometry are sketched in figure~\ref{fig:RYtil-RS3}.  The fivebranes occupy a ball of radius $r_b$.  Outside this ball, a Gaussian angular three-sphere encloses all the fivebranes; the magnetic flux through this sphere backreacts on the geometry to keep the sphere at approximately constant size.  As one dives inside the ball of fivebranes, such a Gaussian sphere encloses less and less fivebrane charge; the flux decreases and the angular $\bS^3$ gradually decreases in size, such that the transverse space smoothly caps off at $r=0$.  On the other hand, the momentum along $\bS^1_\ytil$ carried by the fivebranes exerts pressure, which causes the proper size of the circle to expand in the core of the geometry.  The equation of state for the ensemble geometry is
\begin{equation}
 p=-\rho ~,
\end{equation}
where $p$ is the pressure in the $\tilde{y}$-circle and $\rho$ is the energy density of the source.

As noted in \rcite{Alday:2006nd}, the surface area of the blob is $N^{1/4}$ times larger than the microscopic entropy of the supertube ensemble; equivalently~\rcite{Martinec:2023xvf}, the blob radius $r_b$ is $N^{{1}/{4}}$ times larger than the ``entropy radius'' $\mu$ which would account for the 1/2-BPS entropy $S\sim O(\sqrt{\none\nfive})$ as the horizon area of a small black hole.  Thus, the typical 1/2-BPS state is supported well outside any supposed horizon radius, and so is {\it not} a black hole; rather one should think of such states as {\it fivebrane stars}. 

Note, however, that if one turns off the transverse scalar excitations by hand by making their chemical potential high and the internal mode chemical potential low, the fivebrane windings are brought together.  The distance to the cap grows, as does the coupling there.  The star approaches the black hole threshold, and string perturbation theory breaks down.

The ensemble geometry is spherically symmetric, and that symmetry results in a vanishing of the statistical average $\llangle \sfA \rrangle$ of the one-form $\sfA$.  Similarly, R-R parity forbids an expectation value for the R-R fields $C^{(1)},C^{(3)}$.  Individual elements of the ensemble, however, have expectation values of these fields given by~\eqref{gen STube NS}-\eqref{HpRR}.  While the R-R fields (and the one-form $\sfA$) vanish in the ensemble average, there are substantial statistical fluctuations.  Two-point functions provide an estimate of the RMS average of these quantities.

Consider for instance the fluctuations of $C^{(1)}$ sourced by $\partial_v\mfa^{(0)}$.  The calculation of the torus two-point function of $\sfZ_0$ is much the same as that of $\sfH_5$ apart from the Green's function for $\mfa^{(0)}$ involving $\tau_\parallel$ instead of $\tau_\perp$.  The two-point function of $\sfH_5$ was computed in~\rcite{Raju:2018xue} and found to be of the same order as the square of the one-point function.  Thus we find that the typical element of the ensemble will have RR potentials of the same order as the potential $\sfH_5$, scaled by the internal excitation level $N_\parallel$, even though the ensemble average $\llangle \sfZ_0\rrangle$ vanishes for reasons of symmetry.

For the capped geometry to be reliably weakly-coupled, one wants to check that the likelihood of a self-intersection of the fivebrane profile $\sfF(v)$ is small.  The ensemble averaged amplitude for two points on the profile to be separated an amount $d$ in fact vanishes smoothly as $d\to 0$, see figure~\ref{fig:separation}, and so the typical profile does not have self-intersections (as one would expect for a random walk in five spatial dimensions). 
It thus appears that semiclassical string theory applies to typical 1/2-BPS states, and that quantum fluctuations in the geometry should be small.  Indeed, it was shown in~\rcite{Martinec:2023gte} that the measure of relative quantum fluctuations
\be
\label{state relfluct}
\frac{\int dv \int dv' \langle \sfH_5(\sfx,v)\sfH_5(\sfx',v')\rangle}{\int dv \int dv' \langle \sfH_5(\sfx,v)\rangle\,\langle\sfH_5(\sfx',v')\rangle}
\ee
in the round supertube geometry~\eqref{Fround} is small outside of roughly the 6d Planck scale separation from the fivebrane source.  Again, this is what one expects from the effective string picture~-- fluctuations are controlled by the length scale set by the effective string tension $\mu\sim \ell_{\rm pl}^{\sst (6)}$.  On the other hand, the ensemble average that leads to the ensemble geometry~\eqref{NS5P ensemble} has statistical fluctuations which are large on the scale $r_b$ rather than the scale $\mu$, as we now show.

\subsection{Two-point function of BPS densities}
\label{sec:density 2pt}

A measure of statistical (as opposed to quantum) fluctuations is the density-density correlation in the grand canonical ensemble. The fivebrane charge density is simply an appropriately normalized delta function at its transverse location. The two-point correlation function is denoted by 
\begin{align}
& \llangle ~\!\!\delta^4 (\sfx\tight-{\sfF}(v_1))  \! \;\, \!~\delta^4 (\sfy\tight-{\sfF} (v_2))~\! \!\rrangle ~.
\end{align}
Utilizing the Fourier representation of the delta-function, repeating the analysis of~\rcite{Martinec:2023gte} leads to
\begin{align}
\LLangle\! \!~e^{i \vec{k}_1 \cdot \vec{\sfF}(v_1)}\! \;\, \!e^{i \vec{k}_2 \cdot \vec{\sfF}(v_2)}~\!   \!\RRangle  =e^{-\frac{1}{2} G (0|\tau) (k_1 ^2 + k_2 ^2) -G_{12}(v_{12}|\tau) \vec{k}_1 \cdot \vec{k}_2}~,
\end{align}
where the evaluation of the torus amplitude as a Hilbert space trace over the left-moving, non-zero mode oscillators, with a mode number cutoff $n\leq \Lambda$, yields the Green's function and its coincidence limit
\begin{align}
\label{osc trace}
 G_{12} (v_{12}|\tau) &= \frac{\mu^2}{2} \sum_{n=1} ^{\Lambda} \frac{q^n}{n(1-q^n)} \Big( e^{i n(v_1-v_2)}+e^{i n(v_2-v_1)} \Big)+\frac{\mu^2}{2}\sum_{n=1} ^{\Lambda} \frac{1}{n} e^{in (v_2 -v_1)}~,
\nn\\[.3cm]
 G (0|\tau) &= \mu^2 \sum_{n=1} ^{\Lambda} \frac{q^n}{n(1-q^n)} + \frac{\mu^2}{2} \sum_{n=1} ^{\Lambda} \frac{1}{n}~.
\end{align}
One then has
\begin{align}
\label{dense 2pt}
 &  \llangle ~\!\!\delta^4 (\sfx\tight-{\sfF}(v_1))   \; \delta^4 (\sfy\tight-{\sfF} (v_2))\! \!~\rrangle = \int \frac{d^4 k _1}{(2\pi)^4}\int \frac{d^4 k _2}{(2\pi)^4} e^{- i \vec{k}_1 \cdot \vec{\sfx}} e^{-i \vec{k}_2 \cdot \vec{\sfy}} \LLangle\! \!~e^{i \vec{k}_1 \cdot \vec{\sfF}(v_1)}\! \;\, \!e^{i \vec{k}_2 \cdot \vec{\sfF}(v_2)}\!~  \!\RRangle   
 \nonumber\\[.2cm]
&\hskip 2cm
=\int \frac{d^4 k _1}{(2\pi)^4}\int \frac{d^4 k _2}{(2\pi)^4} e^{- i \vec{k}_1 \cdot \vec{x}} e^{-i \vec{k}_2 \cdot \vec{y}} e^{-\frac{1}{2} G (0|\tau) (k_1 ^2 + k_2 ^2) -G_{12} (v_{12}|\tau) \vec{k}_1 \cdot \vec{k}_2}
\\[.2cm]
&\hskip 2cm
=\frac{1}{(2\pi)^4 (G_{12} (v_{12}|\tau)^2-G^r(0|\tau)^2)^2}  e^{-\frac{G (0|\tau)}{2(G (0|\tau)^2 - G_{12}(v_{12}|\tau)^2)} \big(|\vec{x}|^2+|\vec{y}|^2-2\frac{G_{12}(v_{12}|\tau)}{G (0|\tau)}\vec{x} \cdot \vec{y}\big)}~.
\nonumber
\end{align} 
As a check, one recovers the factorized answer in the formal limit $G_{12}\to 0$, and the integral of the result over space evaluates to $1$.

We want to take a limit $-i\tau\to 0$ and $-i\bar{\tau}\to \infty $ in order to suppress contributions from right-moving oscillator states, while assigning small chemical potential for left-moving oscillator states in order to have large excitation level.     
In this limit,
\begin{align}
\label{Glims}
 G_{12} (v_{12}|\tau) ~\longrightarrow&~ \frac{i}{2\pi  \tau}\frac{\mu^2}{2} \sum_{n=1} ^{\Lambda} \frac{q^n}{n^2} \Big( e^{i n(v_1-v_2)}+e^{i n(v_2-v_1)} \Big)+\frac{\mu^2}{2}\sum_{n=1} ^{\Lambda} \frac{1}{n} e^{in (v_2 -v_1)}~,
\nn\\[.3cm]
 G (0|\tau) =& ~\frac{i\mu^2}{2\pi  \tau} \sum_{n=1} ^{\Lambda} \frac{1}{n^2} + \frac{\mu^2}{2} \sum_{n=1} ^{\Lambda} \frac{1}{n}~.
\end{align}
In this sum, it is understood that both $qe^{iv_{12}}$ and $qe^{-iv_{12}}$ have magnitude less than one, so that the sums in the first term of $G_{12}$ converge when $\Lambda \to \infty$.  This means that as $\Im\,\tau\to0$, $v_{12}$ becomes real, and we see that $\Re\, G_{12}$ is always strictly less than $\Re\, G(0)$.  As a result, the quadratic form in~\eqref{dense 2pt} is negative definite.  Thus we see that the density-density correlator~\eqref{dense 2pt} is a sensibly decaying function of the separation of the observation points.

Now we wish to remove the cutoff.  A naive normal ordering involving the subtraction of $\frac{\mu^2}{2} \sum_{n=1} ^{\Lambda} \frac{1}{n}$ from $G(0|\tau)$, would lead to the pathology that $\Re\, G_{12}(v_{12}|\tau)$ could exceed $\Re\, G(0|\tau)$; however, the effects of the zero-point fluctuations that lead to this issue are vastly subdominant to the effects of the highly excited bath of oscillations at small chemical potential.  We bypass this issue by working in the order of limits $-i\tau\to 0$ first, and then $\Lambda\to\infty$; then the first terms on the RHS's of~\eqref{Glims} vastly exceed the second terms, and so we drop the latter in the limit.
The Green's function then reduces to
\begin{align}
 &G_{12} (v_{12}|\tau) ~\longrightarrow~ \frac{i}{2\pi  \tau}\frac{\mu^2}{2} \Big( Li_{2} \big(e^{i (v_1-v_2)}\big) +Li_{2} \big(e^{i (v_2-v_1)}\big) \Big)-\frac{\mu^2}{2}\log\Big( 1-e^{i(v_2-v_1)}\Big)~.
\end{align}
The identity
\begin{equation}
 Li_2 (z) + Li_2 \big({1}/{z} \big) = -\frac{\pi^2}{6} -\frac{1}{2} \big(\log(-z)\big)^2
\end{equation}
allows the further simplification
\begin{equation}
\label{G12lim}
  G_{12} (v_{12} |\tau) ~\longrightarrow~ \frac{i\mu^2}{12\pi \tau} \Big( \pi^2+\frac{3}{2}v_{12}\big(v_{12}-2\pi \big) \Big) ~,
\end{equation}
where we have taken the branch of the logarithm such that $|\log(-e^{iv_{12}})|\le\pi$, for $0\le v_{12}\le 2\pi$.
Also, $G(0|\tau)$ is given by
\begin{equation}
G (0|\tau) ~\longrightarrow~ \frac{i\pi \mu^2}{12\tau}~,
\end{equation}
up to logarithmic corrections which are subleading in the large $N$ limit.

Alternatively, one can evaluate the Green's function on the torus via path integral methods, with the result \eg\ \cite{Polchinski1}
\begin{equation}
\label{nonchiralGreen}
	G_{12}(\nu_{12}|\tau) = -\frac{\mu^2}{2} \log\big| \theta_1 (\nu_{12}|\tau)\big|^2+\mu^2 \frac{\text{Im}(w_{12})^2}{4\pi \tau_2} +k(\tau,\bar{\tau})~,
\end{equation}
where we set
\begin{equation}
	q=e^{2\pi i \tau}~,~w= 2\pi \nu ~,~ z= e^{-iw}~;
\end{equation}
the additive constant $k(\tau,\bar{\tau})$ is given by
\begin{equation}
	k(\tau,\bar{\tau}) = \frac{\mu^2}{2} \log\left| \eta(q)\right|^2~.
\end{equation}
We also subtract a Casimir energy contribution from the right-movers, $-\frac{\mu^2}{2}\log\left(\bar{q} ^{\frac{1}{12}}\right)$, from Eq.~(\ref{nonchiralGreen}).  Using the modular transformation property of the Jacobi theta function, and taking the limit $-i\tau \to 0, -i\bar{\tau}\to \infty$, we have
\begin{equation}
\label{G12limPI}
G_{12} ~\longrightarrow~
\frac{i\pi \mu^2 \nu^2}{2\tau}-\frac{i\pi\mu^2 \nu}{2\tau}+\frac{i\pi \mu^2 }{12 \tau}-\frac{\mu^2}{2}\log\big(2\sin(\pi \bar{\nu}_{12})\big)~.
\end{equation}
Identifying $\nu = \frac{v_{12}}{2\pi}$ produces an agreement between Eq.~(\ref{G12limPI}) and Eq.~(\ref{G12lim}) (apart from the antiholomorphic contribution, which we had dropped in the evaluation of~\eqref{osc trace}).
Note that the correlations/fluctuations of the left-movers are governed by thermal scale $r_b^2\sim \mu^2/\tau$, while those of the right-movers are governed by the quantum scale $\mu^2$ (the inverse tension of the effective string).

The figure of merit for fivebrane density fluctuations is the connected correlator.  The one-point function of the charge density is given by
\be
\llangle  \delta^4(\sfx-\sfF(v) \rrangle = \frac{1}{(2\pi G (0|\tau))^2 } \, e^{-\frac{|\vec{x}|^2}{2G (0|\tau)}} ~,
\ee
which is consistent with averaging Eq.~(\ref{dilatonresult}) of the Laplacian of $\sfH_5$, where on the LHS one inserts the ensemble average of $\sfH_5$ from Eq.~(\ref{NS5P ensemble}). 
Note that the charge density inside the blob follows from scaling~-- one has a fixed number of fivebranes inside a sphere of radius $N^{1/4}$, \ie\ a coordinate volume of $N$, so the density scales as $N^{-1}$.

Defining the normalized coordinates $\tilde{x} ,\tilde{y} $ and variance ratio $\Delta$
\begin{equation}
\tilde{x} \equiv \frac{\vec{x}}{\sqrt{G (0|\tau)}} 
~~,~~~~ 
\tilde{y} = \frac{\vec{y}}{\sqrt{G(0|\tau)}}
~~,~~~~ 
\Delta = \frac{G_{12}}{G(0|\tau)}
~,
\end{equation}
the connected correlator of density fluctuations is given by
\begin{align}
\label{NormalizedDelta2}
& 
{\llangle \delta^4 (\tilde\sfx\tight-{\tilde\sfF}(v_1))   \; \delta^4 (\tilde\sfy\tight-{\tilde\sfF} (v_2))\rrangle_{\rm conn.}}
\nn\\[.3cm]
&\hskip 2.5cm
=  \frac{1}{(2\pi)^4(1-\Delta^2)^2} \,  e^{-\frac{1}{2(1-\Delta^2)} \big(|\tilde{x}|^2+|\tilde{y}|^2-2\Delta\tilde{x} \cdot \tilde{y}\big)}
- \frac{1}{(2\pi)^4} \,  e^{-\frac{1}{2} \big(|\tilde{x}|^2+|\tilde{y}|^2\big)}
\end{align}
The result typically (\ie\ for generic $v_{12}$) decays exponentially outside the fivebrane blob of radius $r_b$, since there is little source there and hence little to fluctuate.  Inside the blob, since $\Delta$ is of order one, the density fluctuations are of order one.  Also, note that as $v_{12}\to 0$, one has from~\eqref{Glims} that $\Delta\to1$, and hence the variance in the first term in the connected density correlator vanishes, while the prefactor diverges.  This term becomes sharply peaked near $\sfx=\sfy$, and in the limit becomes a delta function of $\sfx-\sfy$.

Because the density fluctuations of the source are of order one, upon convolving with the transverse space Green's function the statistical fluctuations of the harmonic functions in the metric coefficients in the BPS ensemble geometry are of order one, as was calculated in~\rcite{Raju:2018xue} and clarified in~\rcite{Martinec:2023gte}.


\subsection{Two-point function of gravitons}
\label{sec:graviton 2pt}

One of the applications of the effective string picture is that it describes fluctuations about the half-BPS backgrounds and the interactions between bulk perturbations like the graviton and brane fluctuations. In general, the metric, dilaton and magnetic B-field fluctuations are the combined response of the fivebranes together with the bulk geometry in which they are immersed. A gravitational wave propagating in the geometry exemplifies the latter, and the emission or absorption of gravitons by the branes are examples of the former.  Since in the BPS geometries, the profile functions $\sfF,\mfa$ appear in both brane and bulk actions, one cannot wiggle one without wiggling the other, and this correlation complicates the non-BPS response is when both effects are taken into account.

A naive calculation might put this fact aside, and assume that one could treat the absorption/emission of supergravitons from the Polyakov effective string independent of the greybody factor for the propagation of these quanta through the bulk (the result is at least a component of the correct procedure).  Even then, note that the dependence of the brane effective action on the various profile functions $\sfF,\mfa$ is nonlinear, as we saw in section~\ref{sec:gauge profiles}, and so the free field approximation to their dynamics is simply the leading order in a perturbative approximation.

In this approximation, the connected part of the two-point function of supergravitons on the effective string is given by the Wick contraction of supergraviton vertex operators
\begin{align}
	& \LLangle 	e_{1 \mu \nu}  \partial \sfF^{\mu} \bar{\partial} \sfF^{\nu} e^{i \vec{k}_1\cdot \vec{\sfF}(z_1,\bar{z}_1) } \;
 e_{2\alpha \beta}\partial \sfF^{\alpha} \bar{\partial} \sfF^{\beta} e^{i \vec{k}_2\cdot \vec{\sfF}(z_2,\bar{z}_2) } \RRangle_{conn}~,
\end{align}
using the nonchiral torus Green's function Eq.~(\ref{nonchiralGreen}).

The degeneration limit simplifies the expression for the torus Green's function. 
In the limit of $\tau\to i0$, $\bar\tau\to -i\infty$, 
we have~\eqref{G12limPI}, and the derivatives
\begin{align}
\label{Tools}
 \partial_{v_1} G_{12} = \frac{i\mu^2}{4\pi \tau}\big(v_{12}-\pi \big) 	
~~&,~~~~ 
 \partial_{v_1}^2 G_{12} = \frac{i\mu^2}{4\pi \tau} ~~,
\nn\\[.2cm]
	\partial_{u_1} G_{12}=- \frac{\mu^2}{4} \frac{e^{-i \frac{u_{12}}{2}}}{\sin\left( \frac{u_{12}}{2}\right)}
~~&,~~~~
	 \partial_{u_1} ^2 G_{12} = \frac{\mu^2}{8\sin (u_{12} /2)^2 }~.
\end{align}
After some algebra, one finds
\begin{align}
& \LLangle 	e_{1 \mu \nu}  \partial \sfF^{\mu} \bar{\partial} \sfF^{\nu} e^{i \vec{k}_1\cdot \vec{\sfF}(z_1,\bar{z}_1) } \;
e_{2\alpha \beta}\partial \sfF^{\alpha} \bar{\partial} \sfF^{\beta} e^{i \vec{k}_2\cdot \vec{\sfF}(z_2,\bar{z}_2) } \RRangle_{conn}
\nonumber\\[.2cm]
&\hskip 1.5cm 
\sim~   e_{1 \mu \nu}\, e_{2\alpha \beta}\,  \frac{i\mu^4\,e^{-\frac{1}{2} (k_1 ^2 + k_2 ^2)G(0|\tau) - \vec{k}_1 \cdot \vec{k}_2 G_{12}}}{32\pi \tau \sin^2 (u_{12}/2)} \bigg[\delta^{\mu \alpha} \delta ^{\nu \beta} - i\frac{\mu^2 (v_{12}-\pi)^2}{4\pi \tau} k_1 ^{\alpha} k_2 ^{\mu} \delta^{\nu \beta} \nonumber\\ 
&\hskip 5cm
	   +i\frac{\mu^2}{2} e^{-i u_{12}} k_1 ^{\beta} k_2 ^{\nu} \delta^{\alpha \mu}  -\frac{\mu^4 (v_{12} - \pi)^2}{8\pi \tau} e^{i u_{12}}k_1 ^{\alpha} k_1 ^{\beta} k_2 ^{\mu} k_2 ^{\nu} \bigg] ~. 
\end{align}

The radius of the blob and the chemical potential in this limit are related by~\rcite{Martinec:2023xvf}
\begin{equation}
	r_b ^2 = 2G(0|\tau) = \frac{i\pi \mu^2}{6\tau}~,
\end{equation}
and so we have
\begin{align}
\label{2ptGravNOT}
	& \LLangle 	e_{1 \mu \nu} \partial \sfF^{\mu} \bar{\partial} \sfF^{\nu} e^{i \vec{k}_1\cdot \vec{\sfF}(z_1,\bar{z}_1) } \;
 e_{2\alpha \beta} \partial \sfF^{\alpha} \bar{\partial} \sfF^{\beta} e^{i \vec{k}_2\cdot \vec{\sfF}(z_2,\bar{z}_2) } \RRangle_{conn}
 ~\sim ~
\nn \\[.2cm] 
&\hskip 1.5cm
e_{1 \mu \nu}\, e_{2\alpha \beta}\, \frac{3 r_b ^2\mu^2\,e^{-\frac{1}{2} (k_1 ^2 + k_2 ^2)G(0|\tau) - \vec{k}_1 \cdot \vec{k}_2 G_{12}}}{16\pi^2  \sin^2 (u_{12}/2)} \Big[\delta^{\mu \alpha} \delta ^{\nu \beta} - \frac{3 r_b ^2 (v_{12}-\pi)^2}{2\pi^2 } k_1 ^{\alpha} k_2 ^{\mu} \delta^{\nu \beta} \Big] ~.  
\end{align}
Reflecting the structure of the propagator~\eqref{G12limPI}, the result is normalized by the product $\mu r_b$ of the scale of fluctutaions of the left- and right-moving excitations of the fivebrane source.  The fluctuations are of order this scale for momenta up to the inverse blob size, and gaussianly suppressed for larger $k$ due to the smoothness of the fivebrane distribution in the ensemble.

\section{BPS and Near-BPS perturbations}
\label{sec:perts}

The initial derivations of 1/2-BPS NS5-P supertube geometries proceeded by starting with the geometry sourced by an (F1-P) excited string~\rcite{Dabholkar:1995nc}, and then following the solution through a chain of dualities \rcite{Lunin:2001fv,Kanitscheider:2007wq}.  We have rederived these solutions directly in the NS5-P duality frame using the combined action of bulk supergravity action together with that of a fivebrane source.  One of our motivations for doing so is to explore the near-BPS regime; to properly treat such excitations, one must work directly in the appropriate duality frame in order to capture the relevant light degrees of freedom.

\subsection{BPS perturbations}
\label{sec:BPS perts}

Before delving into the near-BPS regime, it is worth pausing to connect the effective action formalism to what we know about perturbations that act within the BPS sector.  The BPS perturbations of the round supertube of figure~\ref{fig:Circular NS5-P} were investigated in~\rcite{Martinec:2020gkv,Martinec:2022okx}.  In the gauged WZW model~\eqref{nullGWZW}, the background is sourced by a brane profile~\eqref{Fround} in which only a single momentum mode of mode number $k$ is populated $N/k$ times.  Half-BPS vertex operators of $\sutwo$ spin $j$ mediate transitions between 1/2-BPS states in which the $2j+1$ background modes are converted into a single mode of momentum $(2j\tight+1)k$, while the polarization of the vertex operator determines the polarization of the resulting mode (for instance, NS-NS vertex operators produce a transverse mode $\sfF$, while R-R vertex operators produce an internal mode $\mfa$).

The supergravity plus fivebrane effective action can be expanded in fluctuations around a solution 
such as~\eqref{NS5-P geom}, \eqref{NS5-P harmfns}, \eg\
\begin{align}
\sfF &= \bar\sfF(v) + \delta\sfF(u,v)
\nn\\
G_{\mu \nu} &= \bar G_{\mu \nu}(v,\sfx) + \delta G_{\mu \nu}(u,v,\sfx)
\end{align}
and similary for the other supergravity fields.  The fivebrane effective action~\eqref{Source} will then contain perturbations such as
\be
\delta\cS = -\half \tau_{\rm\sst NS5} \int \! d^2\sigma \sqrt{-\gamma_2} \,\gamma^{ab}\, e^{-2\bar \Phi(u,v,\bar\sfF+\delta\sfF)} \delta G_{\mu\nu}(u,v,\bar\sfF\tight+\delta \sfF) \,\partial_a(\bar\sfF\tight+\delta\sfF)\,\partial_b(\bar\sfF\tight+\delta\sfF) ~.
\ee
Expanding out in powers of $\delta\sfF$, one finds at $\ell^\th$ order a bulk perturbation involving \eg\ $\ell$ derivatives of $e^{-2\bar\Phi}\delta G$, and  $\ell+2$ powers of $\delta \sfF$.  For symmetric backgrounds such as the round supertube, this derivative expansion can be organized into tensor spherical harmonics.  The 1/2-BPS perturbation that has spin $j=\ell/2$ indeed involves $2j+2$ fluctuations $\delta\sfF$, 
\be
\Big[\partial_{i_1}\cdots\partial_{i_\ell}\big( e^{-2\bar\Phi}\delta G_{jk}\big)\Big]_{\sfx=\bar\sfF}\, \delta\sfF^{i_1}\dots\delta\sfF^{i_\ell}\partial_a\delta\sfF^i\partial_b\delta\sfF^k ~,
\ee
and thus can mediate a transition which annihilates $2j+1$ background modes and creates in their stead a single mode of the requisite quantum numbers.

This structure is analogous to the analysis of gauge theory operators that absorb dilaton partial waves onto a D3-brane~\rcite{Klebanov:1999xv}; and perhaps more to the point, to the operator map in the decoupled fivebrane theory (so-called {\it double-scaled little string theory})~\rcite{Aharony:1998ub,Aharony:2004xn} relating supergraviton vertex operators and their counterparts in the fivebrane gauge theory.
Thus, even in the decoupled theory, one should be able to use the brane effective action to map the perturbations around a particular 1/2-BPS state to excitations of the brane and bulk.  The fact that a single 1/2-BPS perturbation simultaneously deforms the brane source and the ambient bulk geometry is reflected in a property of the string vertex operators in these backgrounds which one can call {\it generalized FZZ duality}~\rcite{Martinec:2020gkv}.  In the exact worldsheet CFT for the round supertube, each vertex operator has a fivebrane component as well as a bulk supergravity component, related by an exact operator identification of the affine $\sltwo$ representation theory involved in the coset model~\eqref{nullGWZW} known as FZZ duality.  In the effective action formalism developed here, this duality is simply the statement that the perturbations $\delta\sfF,\delta\mfa$ of the profile functions appear both in the fivebrane effective action and in the bulk supergravity effective action; one cannot wiggle them independently, but instead only in a correlated fashion.

We will not further develop here the details of the map of 1/2-BPS perturbations in the effective action, leaving the details to future work. 
 One can also consider 1/4-BPS excitations, which add another charged excitation to the background (string winding in an NS5-P background, or momentum in the T-dual NS5-F1 background).  Their features will be briefly described in the discussion section below.

\subsection{Near-BPS perturbations}
\label{sec:near-BPS}

What is the fate of the 1/2-BPS or 1/4-BPS fivebrane star when we excite it away from the BPS bound?  Infinitesimally away from the BPS bound, one might expect a structure analogous to another solitonic system, that of BPS multi-monopoles.  There, the near-BPS dynamics boils down to a sigma model on the multi-monopole moduli space.  The analogous result here would be slow motion on the 1/2-BPS configuration space.  Unless there is some biasing effect that focuses the effective string in a particular direction, it should ballistically and ergodically explore this configuration space.

Occasionally the string will come close to a self-intersection.  The S-dual of this phenomenon~-- intersecting D-branes~-- was analyzed in~\rcite{Douglas:1996yp}.  There, the close approach of two D-branes leads to the excitation of stretched strings between them.  The S-dual would be the excitation of D-branes stretching between nearly coincident fivebranes (separated by the scale $\mu$).  The result would be the trapping of the fivebranes at the intersection.  Thermalization of the excess energy at the intersection leads to the formation of a small 6d black hole, with a pair of supertube strands threading the horizon.

Apart from this sort of rare encounter, when the fivebrane absorbs energy, its gravitational self-attraction will exceed the repulsion caused by the NS $B$-field.  The profile should have a tendency to shrink.  Within the BPS configuration space, the transverse string profile can shrink in either of two ways (in any combination):  First, by diverting more of the excitation budget into excitations of the internal gauge modes $\mfa$; and second, by increasing the average mode number of transverse excitations $\sfF$.  Both of these changes decrease the total amplitude of the transverse profile of the fivebranes. 
Such changes in the excitation profile moves us away from mode equipartition, if the initial state has $\tau_\perp=\tau_\parallel$, and thus results in a reduction in entropy.  Naively, the effect of interaction among the modes would tend to restore equipartition.  Effective strings with a typical mode distribution will simply shrink due to the excess gravitational self-attraction.  
Does the fivebrane star undergo a core-collapse instability, or does the free energy cost of compressing it result in some overall stability against at least small non-BPS excitations?

An ordinary star would compress to a slightly smaller radius, but the susceptibilities of the BPS fivebrane star appear to be large~-- the BPS configurations are pressureless, because supersymmetry entails a cancellation between gravitational, dilatonic and magnetic flux forces in the transverse space.  Thus configurations of vastly different size are energetically degenerate.
Putting more of the effective string excitation budget into internal (gauge) excitations results in a decrease of $r_b$; similarly, if the distribution of transverse modes is biased toward higher mode numbers (longer windings in the NS5-F1 frame), the transverse size of the profile also decreases.  Smaller $r_b$ leads to more strongly peaked harmonic functions $\sfH_5,\sfH_p$, and thus deeper throats with larger redshifts to the cap in the geometry, as well as greater string coupling in the cap.  Thus there are easy routes for the NS5 star to collapse.

One can expand the effective action around the equations of motion around a BPS background
\be
\sfF = \bar\sfF + \delta\sfF  
~~,~~~~
G_{\mu \nu} = \bar G_{\mu \nu} + \delta G_{\mu \nu}
\ee
and similarly for all the other fields.  The solution at zeroth order is a BPS solution of the form written above in section~\ref{sec:NS5 effact}.  At next order we have a free field equation of motion for $\delta\sfF$, which couples to a linear order bulk perturbation.  

We work with the dual variables
\begin{equation}
\label{dualframe}
  \widetilde{G}_{\mu \nu} = G_{\mu \nu} e^{-2\Phi}
~~,~~~~ 
\widetilde{\Phi}=-\Phi 
~~,~~~~
\widetilde{B}_{\mu \nu}~~,
\end{equation}
and denote the dual background field values with a bar, \eg\ $\bar{G}_{\mu \nu} \equiv  G_{\mu \nu} e^{-2\Phi}|_{\text{backgd}}$, $\bar B_{\mu\nu}\equiv\widetilde B_{\mu\nu}|_{\text{backgd}}$, \etc.
The source action can be expanded in fluctuations
\be
\cI = \cI_0+\cI_1+\cI_2+\dots ~.
\ee
At leading order we have the classical action of BPS configurations; the next order vanishes as a result of the BPS equations of motion.  At second order we have
\begin{align}
\label{DeltaI2} 
\cI_2=&-\frac{\tau_{\NSfive}}{4}\int d^2 \sigma \sqrt{-\gamma} \left( \gamma^{ab} \partial_{\alpha}\partial_{\beta} \bar{G}_{\mu \nu}  + \epsilon^{ab} \partial_{\alpha} \partial_{\beta} \bar{B} _{\mu \nu} \right)\partial_a \bar\sfF ^{\mu} \partial_b \bar\sfF^{\nu}\delta \sfF^{\alpha}\delta \sfF^{\beta}
 \nonumber\\[.2cm]
	& 
 -\frac{\tau_{\NSfive}}{2}\int d^2 \sigma \sqrt{-\gamma} \left( \gamma^{ab} \bar{G}_{\mu \nu} + \epsilon^{ab} \bar{B} _{\mu \nu}\right) \partial_a \delta \sfF^{\mu}\, \partial_b \delta \sfF^{\nu}~
\\[.2cm]
	& 
 -\tau_{\NSfive}\int d^2 \sigma \sqrt{-\gamma} \left[ \gamma^{ab} \Big(\delta{G}_{\mu \nu}+\partial_{\alpha}\bar{G}_{\mu \nu}\, \delta \sfF^{\alpha} \Big)+ \epsilon^{ab}\Big( \delta\widetilde{B} _{\mu \nu}+\partial_{\alpha}\bar{B}_{\mu \nu}\, \delta \sfF^{\alpha}\Big)\right]\partial_a \delta \sfF^{\mu}\, \partial_b \bar\sfF^{\nu}
 \nonumber
\end{align} 
Qualitatively, one can think of the lines above as (1) Second-order response of the fivebranes to ``background curvature'', (2) Brane fluctuations, and (3) Bulk/brane absorption or emission. 

To evaluate $\cI_2$, consider a gauge where
\begin{equation}
\bar\sfF ^{v}=v~,~~~
\delta \sfF^{v}=0 ~,
\end{equation}
Solving for $\sfF^u $ in terms of oscillator modes, it is given to a good approximation by $u$, plus thermal fluctuations in the stress energy tensor which are suppressed in the large-$N$ limit. We thus have
\begin{equation}
 \bar\sfF ^u \approx u 
 ~~,~~~~ 
 \delta \sfF^u \approx 0~.
\end{equation}
Evaluating~(\ref{DeltaI2}) in the ``S-dual frame'' on the Lunin-Mathur sollutions
\begin{equation}
\bar{G}_{uv} = \frac{1}{H_5} 
~~,~~~~
\bar{B}_{uv} = -\frac{1}{H_5} 
~~,~~~~
\bar{G} _{vi} = \bar{B}_{vi} = \frac{A_i}{H_5}~,
\end{equation}
one finds after a bit of algebra
\begin{align}
\label{I2formula}
\delta \cI_2=\tau_{\NSfive}\int d^2 \sigma   &\bigg[  \partial_u \delta \sfF^{i}\, \partial_v \delta \sfF^{i}+\Big(  \delta{\widetilde G}_{j v}+  \delta\widetilde{B} _{j v}\Big)\partial_u \delta \sfF^{j}  ~\nonumber\\
& +  \Big( \delta{\widetilde G}_{uj} +  \delta\widetilde{B}_{uj}\Big)\partial_v \delta \sfF^{j}  +\Big(  \delta{\widetilde G}_{j k} + \delta\widetilde{B}_{jk}\Big)\,\partial_u \delta \sfF^{j}\, \partial_v \bar\sfF^{k}~\bigg]~.
\end{align}
This expansion of the fivebrane effective action must be combined with a similar expansion of the bulk supergravity effective action to determine the complete linearized fluctuations around a semiclassical supertube background.



\section{Discussion}
\label{sec:discussion}

The present work is the third installment in a series of investigations of fivebrane dynamics.
In the first installement~\rcite{Martinec:2023xvf}, we showed that the NS5-P duality frame describes the near-source structure of 1/2-BPS fivebranes.  In the second~\rcite{Martinec:2023gte}, we showed that quantum fluctuations of 1/2-BPS states are small, while the statistical fluctuations are substantial.  In this third installment, we have shown how the 1/2-BPS backgrounds solve the supergravity equations of motion coupled to a fivebrane source, and extended our analysis to include the internal gauge excitations of the fivebranes and the bulk R-R fields they source.


In future installments of this series, we will extend these results in a variety of directions:
\begin{itemize}
\item
T-duality to the NS5-F1 frame connects the discussion to $AdS_3/CFT_2$ duality; the fivebrane effective action dualizes to include KK-monopole aspects.
\item
The addition of F1 winding charge leads to horizonless 1/4-BPS NS5-F1-P microstates known as {\it superstrata}; we expect that a picture of generic superstrata, and a corresponding ensemble geometry, will emerge from an analysis of the 1/4-BPS equations of motion.
\item
The collapse of fivebrane stars leads to the onset of the black hole phase, which is dual to the deconfinement transition for the ``little string theory'' that governs non-abelian fivebrane dynamics.  The effective action formalism, extended to include little string dynamics on the fivebranes, could provide a framework for understanding aspects of black hole dynamics and the associated information paradox.
\end{itemize}
Let us mention some aspects of these topics.

\paragraph{NS5-F1 1/2-BPS Fivebrane stars:}
Having a detailed picture of the 1/2-BPS NS5-P solutions, we can consider T-duality to the NS5-F1 frame, and its further $AdS_3$ limit.  The momentum charge $n_p$ is relabeled to $n_1$, \ie\ string winding charge around the T-dual circle, which we parametrize by the coordinate $y$.
Continuing to work in the NS5 decoupling limit~\eqref{decoupling}, the geometry is approximately that of the extremal NS5-F1 solution
\begin{align}
ds^2 &= \Big[ \frac1{\sfH_1}\big( -dt^2+dy^2\big) + dx^2_{\bT^4} \Big]_{||} + \nfive\alpha'\Big( d\rho^2+d\Omega_3^2\Big)_\perp
~~,~~~~
\sfH_1 = 1 + \frac{n_1 (\alpha')^2}{V_4 e^{2\rho}}  
\nn\\[.2cm]
B_2 &= \frac1{\sfH_1} dt\wedge dy + \sfb
~~,~~~~
d\sfb = *_\perp d\sfH_5 = \nfive \alpha' *_{\perp}d(e^{-2\rho})
~~~,~~~~
e^{2\Phi} = \frac{n_5}{e^{2\rho} \sfH_1}
\end{align}
The fivebrane star in this geometry simply replaces the harmonic functions by their ensemble averaged versions~\eqref{NS5P ensemble}, with $\sfH_1=1+\sfH_p$.

The geometries exhibit a crossover at $e^{2\rho}\sim \frac{n_1 (\alpha')^2}{V_4}$.  At larger radius, the constant in $\sfH_1$ dominates; one has a linear dilaton throat, and the proper size of the longitudinal circle $\bS^1_y$ is approximately its asymptotic value $R_y$.  At smaller radius, the constant is unimportant, the string coupling saturates to $\frac{\nfive V_4}{n_1 (\alpha')^2}$ (in particular, there is no longer a strong coupling issue), and the geometry is approximately that of a massless BTZ black hole, with $\bS^1_y$ forming the exponentially growing $AdS_3$ azimuthal direction.  In the further decoupling limit $R_y\to\infty$, the $AdS_3$ geometry fills all of space, and the bulk string theory is dual to a holographic $\it CFT_2$.

\begin{figure}[ht]
\centering
\includegraphics[width=0.55\textwidth]{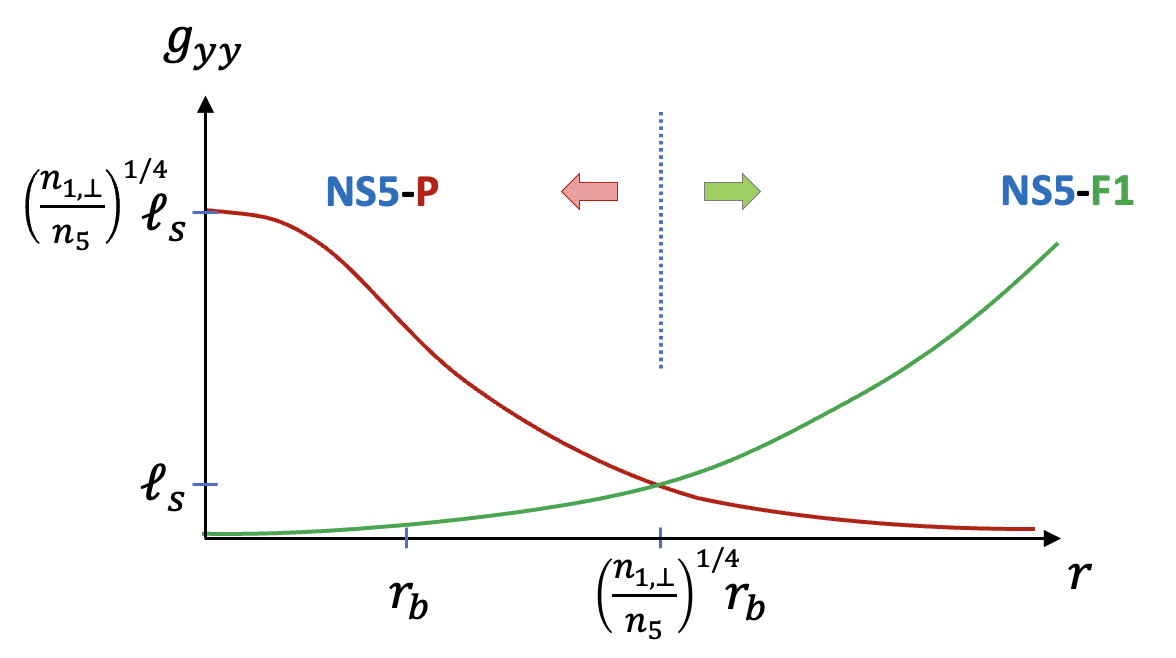}
\caption{ \it
In the $AdS_3$ decoupling limit, the $y$ and $\ytil$-circle sizes as a function of a dimensionless radial coordinate.  At large radius, the NS5-F1 description is appropriate; at small radius, the NS5-P description takes over.
 }
 \label{fig:NS5-F1 Ry}
\end{figure}

However, in this frame the proper size of $\bS^1_y$ falls below the string scale at a radial location 
$(n_1/n_5)^{1/4}$ times larger than the radius $r_b$ of the fivebrane star, see figure~\ref{fig:NS5-F1 Ry}.  At this point the NS5-F1 frame supergravity description breaks down; it would predict a highly curved and cuspy region near the source (the geometry is still capped in the IR, before one gets to a horizon).  This region is better described in the appropriate duality frame for the near-source region, namely the original NS5-P frame, so long as $N_\perp\gg n_5^{\,2}$.

Perturbative string theory knows about T-duality, and so the typical 1/2-BPS state in the NS5-F1 system is not singular; but its microstructure is largely hidden in T-dual NS5-P variables that are invisible to supergravity.  In a forthcoming paper, we discuss the effect of T-duality on the source action.  Depending on whether the circle being T-dualized is parallel or perpendicular to an NS5-brane, it either remains an NS5-brane or becomes a KK-monopole, respectively.  In the present case, the direction being dualized is partly along and partly perpendicular to the brane worldvolume, so the source has both an NS5 and a KKM character.  The brane effective action carries F1 charge via the mechanism of~\rcite{Gregory:1997te}, which is a string-theoretic version of the way a Yang-Mills-Higgs monopole carries electric charge and becomes a dyon (as reviewed for instance in~\rcite{Harvey:1996ur}).
While typical half-BPS states are not properly described by supergravity in this T-dual frame, it is still useful for describing atypical bound states like circular supertubes.

\paragraph{1/4-BPS Fivebrane stars:}

There are nonlinear, fully back-reacted versions of bulk excitations carrying the third BPS charge which are horizonless backgrounds, known as {\it superstrata} (see~\rcite{Shigemori:2020yuo} for a review).  These solutions are still governed by harmonic functions, and the BPS equations can be layered in a hierarchy such that at each order the equations being solved are linear.

The method of superstratum construction begins with an underlying 1/2-BPS supertube (usually this is considered in the NS5-F1 or D5-D1 frame, but there is no reason a priori not to consider the NS5-P frame).  In the first non-trivial step, one modifies the supertube harmonic forms with additional wave modes that vanish at the supertube locus.%
\footnote{The vanishing of the wave excitation at the supertube locus is seen in the explicit solutions constructed to date (see for instance~\rcite{Heidmann:2019zws}), and is inherent in the strategy~-- a non-vanishing wave profile at the underlying fivebrane leads to singularities in the supergravity fields.}
These modified supertube harmonic forms then source the additional harmonic forms activated by the third BPS charge carried by the background.  A key feature is that the modification of the solution amounts to adding a bulk supergravity wave that vanishes at the fivebrane source locus.
As a result, the superstratum deformation does not further excite the underlying fivebranes, and the fivebrane action is largely unaffected, while the bulk action acquires a supergravity wave whose features depend on the brane state. 

While the entropy of supertube solutions scales with the charge quanta as $\sqrt{\nfive\none}$, the entropy of superstrata is estimated to be of order $\sqrt{\nfive\none}\,n_p^{1/4}$~\rcite{Shigemori:2019orj}.  The reason the entropy is not simply the sum of the supertube entropy plus that of a graviton or perturbative string gas, but rather enhanced multiplicatively by a power of the momentum charge, is that the bulk supergravity wave depends on the individual supertube according to the boundary condition that the wave vanishes at the supertube locus.

This interdependence of the superposed wave and the underlying supertube makes it difficult to average over the ensemble of superstratum configurations in the same way one can for supertubes.  
The analysis of 1/4-BPS backgrounds in the effective action, and the properties of generic superstrata, are fertile areas of investigation, which we also hope to explore in future work.

\paragraph{Fivebrane black holes:}

When the transverse excitation profile becomes compact enough, the relative fluctuations~\eqref{state relfluct} become large and the probability of self-intersection becomes substantial.  At this point, the fivebranes merge, and a non-abelian ``little string'' is liberated; we enter the black hole phase.  The ``horizon'' in this effective description is the location where the highly entropic little string phase space can be accessed.

In gauge/gravity duality, the formation of a horizon in the bulk geometry is associated to a deconfinement transition in the dual gauge theory~\rcite{Witten:1998zw}.  In the case of black NS5-branes, the Euclidean geometry associated to this deconfined phase is once again that of the $\frac{\sltwo_\nfive}\uone$  coset conformal field theory,
the ``cigar'' geometry 
\begin{align}
\label{blackNS5}
ds^2 = n_5 \alpha' \big(d\rho^2 + \tanh^2\!\rho\, d\tau^2 \big)
~~,~~~~
e^{2\Phi} = \frac{e^{2\Phi_0}}{\cosh^2\rho}
\end{align}
which thus determines black fivebrane thermodynamics.  The equation of state 
\be
\label{Hag-thermo}
S = \beta\, E
~~,~~~~
\beta =2\pi\sqrt{\nfive\alpha'}
\ee
is the Hagedorn thermodynamics of little string theory~\rcite{Maldacena:1996ya,Berkooz:2000mz,Harmark:2000hw,Kutasov:2000jp}; NS5 dynamics in this regime is dominated by an effective string whose tension scale
\be
\alpha'_{\it little} \equiv \nfive\alpha'
\ee 
is $\nfive$ times lighter than that of the fundamental string.  The heuristic description of these strings is different in different duality frames.  In type IIB, instantons in the 5+1d SYM theory on the fivebranes are solitonic strings; in type IIA, D2-branes stretching between separated NS5-branes are the ``W-strings'' of a 5+1d non-abelian tensor gauge theory that become light when the fivebranes coincide (due to the uncontrolled large coupling that develops).

The black fivebrane geometry~\eqref{blackNS5} generalizes to the superselection sector of $\none$ units of fundamental string winding, which fractionates into $\nfive\none$ units of little string winding (see for instance~\rcite{Maldacena:1996ya}).  
For horizon radius larger than the crossover scale set by the F1 charge, the thermodynamics has approximately the Hagedorn equation of state~\eqref{Hag-thermo}; for horizon radius smaller than this scale, the equation of state is approximately that of BTZ black holes.  One can parametrize the energy as the winding string rest energy for a little string of total winding $\none\nfive$ (corresponding to fundamental string winding $\none$) plus a residual $\vareps$
\be
E = \frac{\none R_y}{\alpha'} + \frac\vareps{R_y}
= \frac{\none\nfive R_y}{\apl} + \frac\vareps{R_y} ~,
\ee
and solve the Virasoro constraints for such a string 
\begin{align}
0 &= -\Big(\frac{\none\nfive R_y}{\apl}+\frac{\vareps}{R_y}\Big)^2 + \Big(\frac{\none\nfive R_y}{\apl} + \frac{n_p}{R_y}\Big)^2 + \frac{4}{\apl}\Big(h-\frac{c}{24}\Big)
\nn\\[.2cm]
0 &= -\Big(\frac{\none\nfive R_y}{\apl}+\frac{\vareps}{R_y}\Big)^2 + \Big(\frac{\none\nfive R_y}{\apl} - \frac{n_p}{R_y}\Big)^2 + \frac{4}{\apl}\Big(\bar h-\frac{c}{24}\Big) ~;
\end{align}
here $(h,\bar h)$ are the left and right excitation levels of the string.
One finds the residual energy and (fractionated) momentum%
\footnote{Not coincidentally, this has the same form as the $T\bar T$ deformation of a CFT (see for instance~\rcite{Cavaglia:2016oda}), with the map of parameters $R_{T\bar T} = 2\pi \none\nfive R_y$ the total length of the little string, and the effective coupling $t_{T\bar T} = \pi\apl$.
}
\begin{align} 
\vareps = -\frac{\none\nfive R_y}{\apl} + \sqrt{\Big(\frac{\none\nfive R_y}{\apl}\Big)^2 + \frac{2}{\apl}\Big(h+\bar h-\frac{c}{12}\Big) + \Big(\frac{n_p}{R_y}\Big)^2}
\quad,\qquad
n_p = \frac{\bar h-h}{\none\nfive}  ~.
\end{align}
The Hagedorn little string entropy 
\be
S_{\it little} \sim
2\pi\Big(\sqrt{h^{\vphantom{|}}}+\sqrt{\bar h}\Big) 
\ee
then reduces to the BTZ entropy in the $AdS_3$ decoupling limit $R_y\to\infty$, 
\begin{align}
\lim_{R_y\to\infty} S_{\it little} &= S_{\rm BTZ} =
2\pi\Big( \sqrt{\none\nfive(\vareps+n_p)/2} + \sqrt{\none\nfive(\vareps-n_p)/2} \,\Big) ~.
\end{align}
It is thus the non-abelian little string regime of the configuration space of fivebranes that accounts for BTZ black hole entropy.

When the fivebranes are separated, the gauge dynamics they support abelianizes; little strings are heavy and decouple (as one sees directly in the type IIA description).  Little string theory governs the strong coupling regime of coincident NS5's, far down the linear dilaton throat.  Black NS5's are associated to the liberation of little strings, much as $AdS_5\times \bS^5$ black holes result from the liberation of the $N^2$ modes of D3 SYM dynamics.

We propose to continue using the effective action upon reaching the collapsed phase, particularly for the near-BPS BTZ regime.  In the spirit of near-susy effective field theory, we need to {\it integrate in} the degrees of freedom which become light -- the BPS Hagedorn little string and light excitations around it.  These are confined to the fivebranes and hence localized where the fivebranes are, in the black hole interior.

According to the fuzzball proposal, the brane gas should not disappear behind the horizon and collapse into a singularity when crossing over into the black hole phase, rather it should describe near-BPS BTZ black holes.  It {\it is} the horizon, as seen from outside.  
The absorption and emission amplitudes of the brane bound state in the weak-coupling regime match those of the corresponding black hole in the strong-coupling regime~\rcite{Callan:1996dv,Dhar:1996vu,Das:1996wn}, at leading order at low energy.  We propose the use of the coupled brane-bulk effective action to extend this agreement to higher energies, and strong coupling, by incorporating some of the brane dynamics into the bulk description.

Bulk+brane actions are of course not new. What is new is the idea to use them in the decoupling limit of gauge/gravity duality, where we are used to thinking that we have to give up one or the other.  We have seen that at least for supertubes (and very likely for superstrata), we can instead keep both in the effective bulk string theory description, with their degrees of freedom in an entangled state so that we are not overcounting degrees of freedom~-- the state of gravitational excitations in the throat of the bulk geometry is strongly correlated to the state of the brane source. 
Our task is to understand the mechanism that maintains that coherence in the black hole phase, and supports the brane structure at the horizon scale, thus allowing the black hole to causally imprint its microstate structure onto the black hole thermal atmosphere, and Hawking radiation to carry away information.  This is the core challenge of the fuzzball program.

An entangled state of brane matter at the horizon scale, and supergravitons in the bulk, is very much the picture one has in the near-extremal limit of rotating BTZ black holes.   The ground state entropy $S_0$ of the left-moving sector of the spacetime CFT is associated to the nonabelian fivebrane dynamics of little strings, as we argued above; while the Schwarzian modes of the near-horizon $AdS_2$ throat of near-extremal rotating BTZ are associated to the vacuum conformal block of the right-movers~\rcite{Ghosh:2019rcj}.  The two are entangled in any particular microstate, and in the ensemble average, in much the same way that we have seen above in the 1/2-BPS sector.

\section*{Acknowledgements} 

We thank 
Iosif Bena,
Ramy Brustein,
Ben Freivogel, 
Jeffrey  Harvey,
Carlos Nunez,
Andrea Puhm,
Erik Verlinde,
and 
Nicholas Warner
for discussions.
The research of YZ is supported by the Blavatnik fellowship at the University of Cambridge.
The work of EJM is supported in part by DOE grant DE-SC0009924. 
EJM thanks the Kavli Institute for Theoretical Physics (KITP) for hospitality during the workshop ``{\it What is string theory?  Weaving perspectives together}'', supported by the grant NSF PHY-2309135.


\vskip 2cm
\appendix
\section{Single source throats}
\label{app:1source}

When there is only a single fivebrane source, the supergravity solutions we have constructed above trivialize.  The reason is that the action~\eqref{GeneralAction} \eg\ for the scalar $\mfa_0$ involves \eg\ $(C^{(1)} + \partial \mfa_0)$, such that a profile $\mfa_0(v)$ can be removed by a gauge transformation of the R-R one-form
\be
\delta C^{(1)} = d\Lambda
~~,~~~~
\delta \mfa_0 = -\Lambda
\ee
that leaves the quantity $\cG^{(1)}$ in the source action~\eqref{GeneralAction} invariant.%
\footnote{The other antisymmetric tensor fields also transform in such a way that the Wess-Zumino term in~\eqref{GeneralAction} and the bulk action are invariant.}
On the other hand, when there is more than one fivebrane, there is no such gauge transformation that will simultaneously remove the scalar excitations on all the fivebranes.

This observation provides another rationale for the absence of throat dynamics for an isolated fivebrane.  To this end, let us add back the asymptotically flat region, and consider the lift of the IIA NS5-P solutions to M-theory, where the transverse scalars $\sfF$ and the internal scalar $\mfa_0$ are treated on an equal footing.  The solution with these fields excited was written in 6d supergravity in~\rcite{Ceplak:2022wri}:
\begin{align}
ds_6^2 & = -\frac{dv}{\sqrt{\sfH_5}}\bigg[ du - 2(1-\sfH_5)\dot\sfF \!\cdot\! d\sfx+\bigg(\dot\sfF^2(v)\,(1-\sfH_5)-\dot\mfa_0(v)^2\Big(1-\frac1{\sfH_5}\Big)\bigg)dv \bigg] + \sqrt{\sfH_5}\,d\sfx\!\cdot\! d\sfx
\nn\\[.3cm]
C^{(1)} &= \dot\mfa_0(v)\Big(1-\frac1{\sfH_5}\Big) dv
~~,~~~~
e^{2\Phi} = g_s^2\sfH_5 ~,
\end{align}
where $\sfH_5$ includes a constant term
\be
\sfH_5 = 1+\frac{\alpha'}{|\sfx-\sfF(v)|^2}
\ee
and overdots denote $v$ derivatives.  The decoupling limit of this solution agrees with the result~\eqref{gen STube NS}-\eqref{HpRR} derived in section~\ref{sec:gauge profiles} above.

Lifting this solution to 11d using the ansatz~\eqref{11d metric}, one finds
\begin{align}
ds_{11}^2 &= e^{-2\Phi/3}\biggl[-du\,dv  - (\sfH_5\tight-1)\Big( 2dv \big(\dot\sfF\!\cdot\! d\sfx \tight+ g_s^2\dot\mfa_0 \, dx_{11} \big) - dv^2\big(\dot\sfF^2\tight+g_s^2\dot\mfa_0^2\big)\Big)  + \sfH_5\big(d\sfx^2 \tight+ g_s^2dx_{11}^2\big) \!+ ds^2_{\bT^4}\biggl]
\nn\\[.3cm]
&= e^{-2\Phi/3}\biggl[-du\,dv+(\sfH_5\tight-1)\Big( \big(d\sfx-\dot\sfF \,dv\big)^2 + g_s^2\big(dx_{11} - \dot\mfa_0\,dv\big)^2\Big) + \big( d\sfx^2+g_s^2dx_{11}^2\big) + ds^2_{\bT^4} \biggl]
\end{align}
where $x_{11}$ has periodicity $2\pi$ (the asymptotic size of the M-theory circle being encoded in the explicit factors of $g_s$ in the metric).
In the scaling limit $g_s\to 0$ with fixed $\hat\sfx=\sfx/g_s,\hat\sfF=\sfF/g_s$, $x_{11}$ and $\mfa_0$,
the metric reduces to
\be
e^{-2\Phi/3}\biggl[-du\,dv + \hat\sfH_5\Big( \big(d\hat\sfx-\dot{\hat\sfF} \,dv\big)^2 + \big(dx_{11} - \dot\mfa_0\,dv\big)^2\Big) + ds^2_{\bT^4} \biggl]
~~,~~~~
\hat\sfH_5 = \frac{\alpha'}{|\hat\sfx-\hat\sfF(v)|^2}  ~.
\ee
It is now manifest that we can eliminate both $\mfa_0$ and $\sfF$ by a coordinate transformation
\be
\label{coordtransf}
x_{11}' = x_{11} - \mfa_0(v)
~~,~~~~
\hat\sfx' = \hat\sfx - \hat\sfF(v) ~,
\ee
and so in the decoupling limit of a single fivebrane source location (and in particular a single fivebrane), there is no bulk dynamics associated to the profiles $\sfF,\mfa_0$.

Before taking the decoupling limit, this gauge transformation does not die away in the asymptotically flat region, and so is not allowed.  What this means is that the dynamics is gauge-trivial down the single-source throat, and lives in the near-source region outside the throat.  This observation dovetails nicely with the discussion in the Introduction, which argued for this result from a worldsheet perspective.

When there are multiple sources present, terms such as $\dot\sfF_m dv$ and $\dot\mfa_{0,m} dv$ for each fivebrane strand ($m=1,...,\nfive$) are multiplied by different source Green's functions $|\sfx-\sfF_m|^{-2}$, and so a single coordinate transformation of the sort~\eqref{coordtransf} cannot remove them all, and there are nontrivial throat excitations.

\vskip 3cm


\bibliographystyle{JHEP}      

\bibliography{fivebranes}


\end{document}

